\documentclass[11pt]{article}

\usepackage[USenglish]{babel} 
\usepackage[T1]{fontenc}
\usepackage[ansinew]{inputenc}
\usepackage{verbatim}

\usepackage{lmodern} 

\usepackage{graphicx} 

\usepackage{amsmath}
\usepackage{amsthm}
\usepackage{amsfonts}

\usepackage[dvipsnames]{xcolor}

\def\a{\alpha}
\def\b{\beta}

\def\d{\delta}
\def\f{\phi}               
\def\g{\gamma}

\def\k{\kappa}                    
\def\l{\lambda}
\def\m{\mu}
\def\n{\nu}
  
\def\p{\pi}                
\def\r{\rho}                                     
\def\s{\sigma}                                   
\def\t{\tau}

\def\z{\zeta}

\def\L{\Lambda}
\def\O{\Omega}

\def\S{\Sigma}

\def\del{\partial}              
\let\a=\alpha \let\b=\beta \let\g=\gamma \let\d=\delta 
\let\z=\zeta    \let\k=\kappa
\let\l=\lambda \let\m=\mu \let\n=\nu  \let\r=\rho
\let\s=\sigma \let\t=\tau  \let\f=\phi  
    \let\L=\Lambda
    
\let\ep=\epsilon

\def\nn{\nonumber} \def\bd{\begin{document}} \def\ed{\end{document}}
\def\ds{\documentstyle} \let\fr=\frac \let\bl=\bigl \let\br=\bigr
\let\Br=\Bigr \let\Bl=\Bigl
\let\bm=\bibitem
\let\na=\nabla
\let\pa=\partial \let\ov=\overline
\newcommand{\be}{\begin{equation}}
\newcommand{\ee}{\end{equation}}
\def\ba{\begin{array}}
\def\ea{\end{array}}
\def\ft#1#2{{\textstyle{{\scriptstyle #1}\over {\scriptstyle #2}}}}
\def\fft#1#2{{#1 \over #2}}
\def\del{\partial}
\def\sst#1{{\scriptscriptstyle #1}}
 \def\oneone{\rlap 1\mkern4mu{\rm l}}
\def\ie{{\it i.e.\ }}
\def\via{{\it via}}
\def\semi{{\ltimes}}
\def\str{{\rm str}}
\def\tr{{\rm tr}}
\def\Dm{{{D_{\sst{max}}}}}
\def\vac{ \left | 0 \right \rangle }
\def\kvac{ \left | k \right \rangle }

\def\half{\frac{1}{2}}
\def\sp{\; \; \;}

\def\bol{ \left | B (p^+) \right \rangle}
\def\bo1{ \left | B^0 (p^+) \right \rangle}

\def\bolt{ \left | B (p^+) \right \rangle_{\t}}

\def\boxl{ \left | B (x^-) \right \rangle}
\newcommand{\bea}{\begin{eqnarray}}
\newcommand{\eea}{\end{eqnarray}}

\def\<{ \langle }
\def\>{ \rangle }

\def\S{\Sigma}
\renewcommand{\floatpagefraction}{0.6}
\renewcommand{\textfraction}{0.2}

\newcommand\ca{\mathcal{A}}
\newcommand\vp{\varphi}
\newcommand\beal{\begin{align}}
\newcommand\bbone{\ensuremath{\mathbbm{1}}}

\newcommand{\eq}[1]{\begin{equation}#1\end{equation}}
\newcommand{\spl}[1]{\begin{split}#1\end{split}}
\newcommand{\al}[1]{\begin{align}#1\end{align}}
\newcommand{\subeq}[1]{\begin{subequations}#1\end{subequations}}

\newcommand{\arXividhepth}[1]{\href{http://arxiv.org/abs/#1}arXiv:{\tt #1} [hep-th]}
\newcommand{\arXividother}[2]{\href{http://arxiv.org/abs/#1}arXiv:{\tt #1} [#2]}

\newcommand{\bg}[1]{\hat{#1}}
\newcommand{\wj}{\widetilde{J}}
\newcommand{\reo}{\mathrm{Re}~\!\omega}
\newcommand{\imo}{\mathrm{Im}~\!\omega}
\newcommand{\ads}{AdS_4}
\newcommand{\mcal}{\mathcal{M}}
\newcommand{\ccal}{\mathcal{C}}
\newcommand{\ncal}{\mathcal{N}}
\newcommand{\boxedeq}[1]{
\begin{equation}
\fbox{
\rule[0.7cm]{0pt}{0pt}
$#1$
\rule[-0.45cm]{0pt}{0pt}
}
\end{equation}
}

\def\d{\text{d}}

\def\slashchar#1{\setbox0=\hbox{$#1$}           
\dimen0=\wd0                                 
\setbox1=\hbox{/} \dimen1=\wd1               
\ifdim\dimen0>\dimen1                        
\rlap{\hbox to \dimen0{\hfil/\hfil}}      
#1                                        
\else                                        
\rlap{\hbox to \dimen1{\hfil$#1$\hfil}}   
/                                         
\fi}

\def\Re           {{\rm Re\hskip0.1em}}
\def\Im           {{\rm Im\hskip0.1em}}

\newcommand{\E}{\text{\tiny E}}

\newcommand{\ams}{{\it Institute for Theoretical Physics,
University of Amsterdam, \\
Science Park 904, Postbus 94485, 1090 GL Amsterdam, The Netherlands} \\
{\tt J.Smolic@uva.nl, M.Smolic@uva.nl, K.Skenderis@uva.nl, M.Taylor@uva.nl}}
\newcommand{\auth}{{\large B. Gout\'eraux, J.Smolic, M.Smolic, K.Skenderis and M.Taylor}}

\begin{document}
\begin{titlepage}
\begin{center}


\vskip 2cm
{\Large \bf Higher derivative effects for 4d AdS gravity}


\vskip 1.25 cm {\bf Jelena Smolic$^a$ and Marika Taylor$^{a,b}$}
\\ {\vskip 0.5cm \it\small
Institute for Theoretical Physics$^a$, \\
Science Park 904, 1090 GL Amsterdam, the Netherlands. \\
\vskip 0.2cm
School of Mathematics$^b$, \\
Highfield, University of Southampton, UK.
}

{\vskip 0.2cm \small {\it E-mail: J.Smolic@uva.nl and  M.M.Taylor@soton.ac.uk} }

\end{center}

\vskip 1 cm

\begin{abstract}
\baselineskip=16pt
Motivated by holography we explore higher derivative corrections to four-dimensional Anti-de Sitter (AdS) gravity. We point out that in such a theory the variational problem is generically not well-posed given only a boundary condition for the metric. However, when one evaluates the higher derivative terms perturbatively on a leading order Einstein solution, the equations of motion are always second order and therefore the variational problem indeed requires only a boundary condition for the metric. The equations of motion required to compute the spectrum around the corrected background are still generically higher order, with the additional boundary conditions being associated with new operators in the dual conformal field theory. We discuss which higher derivative curvature invariants are expected to arise in the four-dimensional action from a top-down perspective and compute the corrections to planar AdS black holes and to the spectrum around AdS in various cases. Requiring that the dual theory is unitary strongly constrains the higher derivative terms in the action, as the operators associated with the extra boundary conditions generically have complex conformal dimensions and non-positive norms. 

\end{abstract}

\end{titlepage}

\setcounter{tocdepth}{2}

\tableofcontents
\pagebreak


\section{Introduction}

In this paper we will explore higher derivative corrections to
gravity theories in $(3+1)$-dimensions with negative cosmological constant.
Our main motivation for looking at higher derivative corrections to four dimensional AdS black holes is
in the context of holography and, in particular, applied holography, AdS/CMT, where many of the systems of interest are modelled by four dimensional bulk spacetimes.  The addition of higher derivative terms allows us to probe the dual physics as one moves away from infinite $N$ and infinite 't Hooft coupling.

Finite $N$ effects can change the physics qualitatively. For example, let us consider
holographic superconductors, a subject which has been extensively studied in recent years,
initiated by \cite{HoloCMT1}, \cite{HoloCMT2} and  \cite{HoloCMT4}. Working with classical gravity there is an
apparent  violation of the Coleman-Mermin-Wagner theorem. This well-known theorem states that, for system in two spatial dimensions, we cannot have continuous symmetry breaking in systems at finite temperature and hence the formation of a symmetric breaking condensate is forbidden. However, holographic superfluids have been found in $(3+1)$ bulk dimensions, in which a symmetry breaking operator in the dual $(2+1)$ dimensional CFT acquires an expectation value. As explored in \cite{Anninos:2010sq}, this is an infinite $N$ effect and at finite $N$ quantum effects in the bulk indeed ensure that the symmetry breaking operator does not have a well defined expectation value, in accordance with the expected field theory behaviour \cite{Witten:1978qu}.

One does not see a qualitative finite $N$ effect such as the restoration of the Coleman-Mermin Wagner theorem by evaluating higher derivative corrections on the leading order gravity solution but rather by exploring quantum effects in the bulk. Evaluating higher derivative corrections rather shifts the saddle point and allows one to compute corrections to thermodynamic quantities, transport coefficients and so on. In the context of five bulk dimensions, a considerable effort has been put into investigating higher derivative corrections and exploring the effects on the ratio of the shear viscosity $\eta$ to the entropy density $s$, see for example \cite{Buchel:2004di,Benincasa:2005qc,Kats:2007mq,Brigante:2007nu}.

In particular, \cite{Brigante:2007nu} used Gauss-Bonnet curvature corrections and initiated a bottom up exploration of the constraints on the higher derivative corrections imposed by unitarity of the dual CFT.  Working with the Gauss Bonnet term is particularly convenient because the corrections to AdS planar black holes are known analytically for any value of the Gauss Bonnet coupling constant, see \cite{Cai:2001dz}
and also \cite{Nojiri:2001aj, Cho:2002hq,Neupane:2002bf}. Note that an effect of the Gauss-Bonnet term relevant to the superfluids mentioned above was discussed in \cite{Gregory}, where it was found that addition of the higher curvature terms makes condensation to a superfluid phase more difficult.

The Gauss-Bonnet terms, and corresponding corrected AdS black holes, are a useful way to go beyond classical gravity in bulk dimensions higher than four. However, such terms are trivial in four bulk dimensions, in the sense that an Einstein metric is uncorrected and therefore one needs to include higher order curvature invariants to obtain non trivial corrections\footnote{A review of higher order gravity theories and their black hole solutions may be found in \cite{Charmousis:2008kc}.}. An alternative possibility is to couple Einstein gravity to a dilaton in four dimensions because Gauss-Bonnet type corrections to diatonic black holes are then non trivial, see for example \cite{Cai:2009zv}, but this does not address the question of how AdS black holes with no dilaton are corrected.

Apart from AdS/CMT motivations mentioned above, for which dilatonic AdS black holes may indeed already capture many relevant features, see for example \cite{Charmousis:2012dw}, there are a number of other important motivations in exploring higher derivative corrections to Einstein gravity with a negative cosmological constant. The first is in understanding the AdS/CFT correspondence when the dual theory is on an $S^3$.
In recent years there has been considerable progress in understanding dual (supersymmetric) 3d CFTs, following the works of BLG \cite{BLG} and ABJM \cite{ABJM},  and localisation techniques have been used to compute free energies of the dual theories placed on an $S^3$. Taking the limit of large $N$ and large 't Hooft coupling, the free energies have been matched to the onshell renormalised action of $AdS_4$ with an $S^3$ boundary in Einstein gravity \cite{Drukker:2010nc}. Localisation techniques also allow us to access the subleading terms in the free energy which should be compared to the effects of higher derivative terms evaluated on the bulk $AdS_4$. Comparing these subleading terms with the gravity results we develop here can be used to test the correspondence and indeed restrict which higher derivative terms can arise in the four dimensional bulk action.

The second motivation in exploring higher derivative terms in four dimensions is in the context of understanding the holographic dictionary. One of the main points of this paper is that the addition of higher derivative terms generically involves additional data being required for the variational problem to be well-defined. In the context of holography, the additional data corresponds to a new operator in the dual CFT, in addition to the stress energy tensor which is dual to the bulk metric. For generic higher derivative terms added to the action the dual operator has complex dimension and/or negative norm, reflecting the fact that the corrected added violates unitarity. This analysis provides a very direct probe of the unitarity properties arising from the higher derivative terms.

Historically the main context in which higher curvature corrections to four-dimensional gravity has been studied is as a toy model for a quantum theory of gravity. In this context the key problem is that
adding higher curvature corrections adds higher-order time derivatives to the theory and consequently ghosts.
Recently there has been considerable interest in so-called critical gravity theories, in which ghostlike modes appear to be absent, in both three \cite{ChiralTMG} and four \cite{Pope1,Pope2} bulk dimensions.
The four dimensional story that we develop here is the exact analogue of the discussions in
 \cite{Skenderis:2009nt, Skenderis:2009kd} for topologically massive gravity in three dimensions \cite{Deser1,Deser2}: the higher derivative terms in TMG were shown to be associated with a new operator in the dual two dimensional CFT. In TMG, regardless of the value of the coupling of the higher derivative term a violation of unitarity was found in the dual field theory, either by a complex operator dimension or by an operator whose two point function was non-positive. Note that this violation of unitary persisted even at the so-called critical point, where the new operator together with the stress energy tensor were non-diagonalizable.
In this paper we will show that analogous problems are found in the four-dimensional higher derivative theories.

Given that higher derivative terms generically give rise to new boundary conditions and hence dual CFT operators, whose properties are not consistent with unitary,  one may ask how this observation can be consistent with the fact that top down models arising from string theory are necessarily unitary. To understand this point, one should first note that in the context of string theory and holography the higher curvature terms are always viewed as an infinite series of
small corrections to the leading order effective action. The action with higher derivative terms is not quantized, which makes
the issue of ghostlike modes moot. In other words, the effective action takes the form
\be
I = \frac{1}{2 \kappa^2} \int d^4 x \  \sqrt{-g} (R - 2\L + \a_n l_p^n R^n + \cdots )
\ee
where $\Lambda$ is the cosmological constant; $R^n$ denotes schematically an $n$-th order invariant\footnote{Derivatives of the curvature can also arise but will not be considered here.}; $\a_n$ is a dimensionless numerical constant and $l_p$ denotes the effective Planck length. The effective Newton constant in the Einstein theory is $\kappa^2 = 8 \pi G$.

One is by assumption working in a
regime where $l_p$ is small and therefore the corrections should be treated perturbatively. Suppose ${g}_{(0)}$ is a solution of the
Einstein theory, namely
\be
{\cal G}_{\m\n} ({g_{(0)}}) = R_{\m\n}({g}_{(0)}) - \frac{1}{2} R(g_{(0)}) {g}_{(0) \m\n} + \L {g}_{(0)\m\n}
\ee
Then the corresponding solution of the corrected theory can be expressed as a perturbative series
\be
g = g_{(0)} + l_p^{n+1}   g_{(n)} + \cdots
\ee
with
\be
{\cal G}_{\m\n} (g_{(n)}) =  - \a_n \frac{\delta R^n}{\delta g^{\mu \nu}} (g_{(0)})
\ee
and so on.

One should emphasize at this point the conceptual difference between evaluating the higher derivative terms on the lowest order solution and treating the higher derivative term non perturbatively. In the former case,
the equations for all the metric corrections $g_{(n)}$ are second order inhomogeneous differential equations, rather than higher order differential
equations. Since the equations are second order, the only boundary data that needs to be supplied for the variational problem to be well-defined is
the metric. When one is considering the higher derivative terms evaluated on the lowest order solution, an analogue of the Gibbons-Hawking-York \cite{Gibbons:1976ue, York:1973ia} term
in the action can always be defined such that the variational problem is well-defined for a Dirichlet condition on the metric.

By contrast, as we will explore in sections \ref{two} and \ref{spectrum}, whenever the higher derivative terms are treated non-perturbatively or when we consider the spectrum
around a given corrected background, the resulting equations of motion are generically higher order\footnote{The Lovelock theories \cite{Lovelock:1971yv} are a well-known counterexample in which the equations of motion remain second order.}. This means that additional boundary data needs to be supplied. In the context of holography one can understand the additional data as corresponding to additional dual operators in the field theory, beyond the stress energy tensor. The variational problem
in such contexts will be well-defined only if one supplies additional information together with the Dirichlet condition on the metric; the actual information which
is needed depends on which higher derivative terms are added.

Thus, given a background which solves the supergravity equations at leading order, the variational problem will be well-defined when one computes the corrections to this solution without specifying additional data. However, when one looks at the spectrum around this background, the higher order nature of the field equations manifests itself and additional data, corresponding to a new dual operator, is required.

A top down model arising from string theory must be consistent with unitarity. This is guaranteed if the curvature invariant is such that the resulting equations are actually second order. (Note that since one is treating the corrections perturbatively it is guaranteed that the shift to $\eta/s$ is small and is consistent with unitarity, in contrast to the discussions of \cite{Kats:2007mq,Brigante:2007nu}
 in which the coupling constant of the higher derivative term is allowed to be of order one.)

As we discuss at the end of section \ref{spectrum} another case in which the higher derivative invariant is automatically consistent with unitarity is when the linearised field equation around AdS remains second order. This is a weaker condition than requiring that the equation of motion is always second order, but suffices to ensure that there is no non-unitary dual operator induced by adding the higher derivative term. Examples of such curvature invariants are those built out of the Weyl tensor of order three and higher.

Finally it is interesting to note that reducing a curvature invariant of a given order from ten or eleven dimensions to four dimensions on a curved manifold gives rise to curvature invariants in the effective four dimensional action which is both of the same order and of a lower order, see for example \eqref{mixing}. In the context of AdS solutions the reduction required is indeed always on curved manifolds such as spheres. This implies in particular that a curvature invariant such as one quartic in the Riemann tensor never arises without an accompanying term quadratic in the Riemann tensor and a shift of the cosmological constant. Here we show that the term quadratic in the Riemann tensor gives rise to a new boundary condition for the linearised theory around AdS, and hence a dual operator in the CFT, which turns out to be non-unitary. When one combines all terms arising from the corrections at a given order in the upstairs theory, the resulting four dimensional theory must be unitarity and this may be achieved either by the linearised theory around AdS being second order or by the higher order terms conspiring to give a unitary dual operator.

The plan of this paper is as follows. In section \ref{two} we discuss in more detail the variational problem in higher derivative theories and show that it is well-posed with only boundary data for the metric when one treats higher derivative terms perturbatively about a leading order Einstein solution. In section \ref{three} we first discuss what curvature invariants are expected to arise in the effective four-dimensional action from a top down perspective and then we explore the effects of various curvature invariants on four dimensional planar AdS black holes. Our goal is to find an analogue of the Gauss-Bonnet corrected black hole in five and higher dimensions, i.e. a representative corrected $AdS_4$ planar black hole, and we find that the solution in the Weyl corrected theory is the closest analogue. In section \ref{spectrum} we look in detail at the spectrum in theories with curvature squared corrections, demonstrating that there are indeed new dual operators associated with the higher derivative terms and these are non-unitary. Noting that the spectrum in the Weyl cubed theory is unchanged again this seems to be the simplest case of a representative correction. In section \ref{conc} we conclude.

\section{The variational problem in higher derivative theories} \label{two}

In general one cannot define an analogue of the Gibbons-Hawking-York term \cite{Gibbons:1976ue, York:1973ia} such that the variational problem is well-defined with only a Dirichlet condition on the metric - one must impose additional conditions. This observation explains a long standing problem in the literature: for generic higher derivative corrections the analogue of the Gibbons-Hawking-York (GHY) term has never been found.

There is considerable literature discussing the variational problem in higher derivative theories. In the context of corrections arising in string theory, boundary terms were discussed in \cite{Myers:1987yn} where the analogue of the GHY term was found for Gauss-Bonnet. This is a very special case, however, as the field equations are second order. For corrections involving powers of the Ricci scalar, the variational problem was discussed in \cite{Hawking:1984ph}. The generic issues in setting up a variational problem for higher derivative gravity given only a boundary condition on the metric were highlighted in \cite{Madsen:1989rz}: the boundary terms which arise in varying the bulk action cannot in general be integrated to give an analogue of the GHY term.

Here we argue that the problem in finding a GHY term results from the fact that in general such a term {\it cannot exist}: one must specify additional data together with the metric. In special cases an analogue of the GHY term was found, for example, for Lovelock theories. However, Lovelock theories are themselves special in that the equations of motion are actually second order and this fact explains why a GHY term could be found.

A useful approach to dealing with higher derivative theories is the auxiliary field method and the variational problem in such a context was discussed in \cite{Hohm:2010jc}. In this approach the higher order equations are reduced to coupled second order equations for the metric and the auxiliary fields, and 
one specifies boundary data for both the metric and for the auxiliary field. In the context of perturbatively evaluating higher derivative corrections on leading order Einstein solutions, the boundary condition for the auxiliary field does not involve new data, but rather can be built out of the boundary data for the metric. When one looks at the spectrum, however, one sees that there is indeed generically new data required for the auxiliary field. These points will be illustrated further when we use the auxiliary field method to discuss the spectrum in section \ref{spectrum}.

Before moving on to consider specific models for higher derivative corrections in four dimensions, let us discuss the issue with the variational problem. We consider a general action in $(d+1)$ dimensions
\be
I = \int_{\cal M} d^{d+1}x \sqrt{-g} {\cal L},
\ee
where the Lagrangian ${\cal L}$ depends only on the metric and the Riemann tensor. The variation of the action with respect to the metric gives
\bea
\delta I &=& \int_{\cal M} d^{d+1}x \sqrt{-g} \left ( \frac{1}{2} g^{\mu \nu} {\cal L} + {\cal L}^{\mu \nu} \right ) \delta g_{\mu \nu} \\
&& + \int_{\cal M} d^{d+1}x \sqrt{-g} \left ( {\cal L}^{\mu \nu \rho \sigma} R_{\mu \nu \rho}^{\; \; \; \;\; \;  \lambda} \delta g_{\sigma \lambda}
+ 2 \nabla_{\rho} \nabla_{\mu} {\cal L}^{\mu \nu \rho \sigma} \delta g_{\nu \sigma} \right ) \nn \\
&& + 2 \int_{\partial \cal M} d \Sigma^{\mu} {\cal L}_{\mu \nu \rho \sigma} \nabla^{\rho} \delta g^{\sigma \nu} + \cdots \nn
\eea
Here $\partial {\cal M}$ is the boundary of the manifold ${\cal M}$ and we define
\be
{\cal L}^{\mu \nu} = \frac{\delta {\cal L}}{\delta g^{\mu \nu}} \qquad
{\cal L}^{\mu \nu \rho \sigma} = \frac{\delta {\cal L}}{\delta R^{\mu \nu \rho \sigma}}
\ee
while the ellipses denote boundary terms which vanish with a Dirichlet boundary condition on the metric, $\delta g = 0$.

In the case of Einstein gravity
\be
{\cal L} = \frac{1}{2 \kappa^2} (R - 2 \Lambda)
\ee
and thus the boundary term in the variation is
\be
\int_{\partial \cal M} d \Sigma^{\mu} (g^{\nu \sigma} \nabla_{\mu} \delta g_{\nu \sigma} - g_{\mu \sigma} \nabla_{\nu} \delta g^{\nu \sigma} ).
\ee
As is well-known, one can set up a well-defined variational problem by noticing that
\be
\delta \left ( - \frac{1}{\kappa^2} \int_{\partial M} dx \sqrt{-\gamma} K \right ) = - \frac{1}{2 \kappa^2} \int_{\partial M} d \Sigma^{\mu} (g^{\nu \sigma} \nabla_{\mu} \delta g_{\nu \sigma} + \cdots )
\ee
where the ellipses again denote terms which depend only the on restriction of the metric variation to the boundary (and which hence vanish given the boundary condition). Thus the addition of this term, the Gibbons-Hawking-York term, to the action gives a well-defined variational problem in which the metric on the boundary is held fixed.

For generic Lagrangians involving higher powers of the curvature, the boundary terms involving metric derivatives cannot be canceled by those in the variation of a boundary term. To illustrate this it is useful to look at a specific example,
\be
{\cal L} = \frac{1}{2 \kappa^2} (R - 2 \Lambda - \alpha R_{\mu \nu \rho \sigma} R^{\mu \nu \rho \sigma}).
\ee
When $\alpha = 0$ this reduces to the Einstein theory. When $\alpha$ is small the higher derivative term can be treated perturbatively. It is thus useful to express
the equations of motion in the form
\bea
R_{\mu \nu} &=& \bar{T}_{\mu \nu} \equiv \Lambda g_{\mu \nu} + \alpha t_{\mu \nu}; \\
t_{\mu \nu} &=& \frac{1}{(d-1)} R^{\rho \sigma \tau \eta} R_{\rho \sigma \tau \eta} g_{\mu \nu} + \frac{4}{(d-1)} \nabla_{\rho} \nabla_{\sigma} R^{\rho \sigma} g_{\mu \nu} \nn \\
&& - 2 R_{\mu \rho \sigma \lambda} R_{\nu}^{\; \; \rho \sigma \lambda} - 4  \nabla^{\rho} \nabla^{\sigma} R_{\rho \mu \nu \sigma}, \nn
\eea
with $\bar{T}$ being the effective (trace adjusted) stress energy tensor. In later sections we will be interested in four dimensional models, in which the Riemann squared term can be rewritten in terms of the Ricci tensor and the Ricci scalar, but in this section we will work in general dimension. The reason for consider terms involving the Riemann tensor is that such terms will always arise from top down models, and (unlike Ricci scalar and Ricci terms) they cannot be removed by field redefinitions. We can use the Bianchi identities to simplify the stress tensor as
\bea
t_{\mu \nu} &=& \frac{1}{(d-1)} R^{\rho \sigma \tau \eta} R_{\rho \sigma \tau \eta} g_{\mu \nu} - 2 R_{\mu \rho \sigma \lambda} R_{\nu}^{\; \; \rho \sigma \lambda} \\
&& + \frac{2}{(d-1)} \Box R g_{\mu \nu}
- 4  \Box R_{\mu \nu} + 4 \nabla^{\rho} \nabla_{\mu} R_{\nu \rho}, \nn
\eea
where $\Box = \nabla^{\rho} \nabla_{\rho}$.

A perturbative treatment of the field equations means that one looks for a solution such that
\be
g_{\mu \nu} = g_{(0) \mu \nu} + \alpha g_{(1) \mu \nu} + \cdots
\ee
where $g_{(0)}$ is Einstein with cosmological constant $\Lambda$ and $g_{(1)}$ satisfies
\bea
({\cal L}_{R} - \Lambda) g_{(1) \mu \nu} &=& \frac{1}{(d-1)} R^{\rho \sigma \tau \eta}(g_{(0)})
R_{\rho \sigma \tau \eta}(g_{(0)})  g_{(0)\mu \nu} \\
&& - 2 R_{\mu \rho \sigma \lambda} (g_{(0)})
R_{\nu}^{\; \; \rho \sigma \lambda} (g_{(0)}), \nn
\eea
where ${\cal L}_R$ is the linearized Ricci operator and terms on the right hand side are evaluated on the metric $g_{(0)}$ using the connection of that metric. Note that the terms involve derivatives of the Ricci tensor do not contribute since the covariant derivative of the Einstein metric $g_{(0)}$ is zero.
As emphasised earlier, this equation is a second order inhomogeneous equation for $g_{(1)}$ and therefore it does not require any new boundary condition. Note that we regard here the boundary conditions for the metric as being given as a power series in $\alpha$; i.e. the homogenous part of the solution $g_{(1)}$ is determined by this data.

Let us now turn to the question of the variational problem for such a theory.
The new (relative to Einstein gravity) boundary term that arises in varying the action is then
\be \label{new-var}
\frac{2 \alpha}{\kappa^2} \int_{\partial {\cal M}} d\Sigma^{\mu} R_{\mu \nu \rho \sigma} \nabla^{\rho} \delta g^{\sigma \nu} + \cdots
\ee
where we again suppress terms which vanish for the boundary condition $\delta g = 0$. This term can be manipulated using the Gauss-Codazzi relations as follows. The metric on ${\cal M}$ can be decomposed as
\be
ds^2 = (N^2 + N_{\mu} N^{\mu}) dr^2 + 2 N_{\mu} dx^{\mu} dr + \gamma_{\mu \nu} dx^{\mu} dx^{\nu}
\ee
in terms of hypersurfaces $\Sigma_r$ of constant $r$ with the unit normal to each hypersurface being given by $n^{\mu}$. As the notation suggests, we are most interested in the case where the finite boundary is at spatial infinity, so $r$ is indeed a radial coordinate\footnote{Such a foliation would also be appropriate near timelike infinity, in which case $r$ would be a time coordinate and $g_{rr} < 0$ in Lorentzian signature.}.
Defining the radial flow vector $r^{\mu}$ such that $r^{\mu} \pa_{\mu} r = 1$, the components of $r^{\mu}$ tangent and normal to the hypersurfaces define the shift $N^{\mu}$ and the lapse $N n^{\mu}$ respectively. The extrinsic curvature $K_{\mu \nu}$ of the hypersurface is given by
\be
K_{\mu \nu} = \frac{1}{2} {\cal L}_{n} \gamma_{\mu \nu},
\ee
where ${\cal L}$ is the Lie derivative. The Riemann tensor of the $(d+1)$ dimensional manifold can now be expressed entirely in terms of the intrinsic curvature and extrinsic curvature of $\Sigma_r$ via the Gauss-Codazzi relations
\bea
\gamma^{\alpha}_{\mu} \gamma^{\beta}_{\nu} \gamma^{\gamma}_{\rho} \gamma^{\delta}_{\sigma} R_{\alpha \beta \gamma \delta}
&=& \hat{R}_{\mu \nu \rho \sigma} + K_{\mu \sigma} K_{\nu \rho} - K_{\mu \rho} K_{\nu \sigma}; \\
\gamma^{\rho}_{\nu} n^{\sigma} R_{\rho \sigma} &=& D_{\mu} K^{\mu}_{\nu} - D_{\nu} K^{\mu}_{\mu}; \nn \\
n^{\rho} n^{\sigma} R_{\mu \rho \nu \sigma} &=& - n^{\rho} \nabla_{\rho} K_{\mu \nu} - K_{\mu \rho} K^{\rho}_{\nu}, \nn
\eea
where $D_{\mu}$ is the covariant derivative of the metric $\gamma$ and $\hat{R}$ denotes the curvature of this metric. A useful manipulation of these equations gives
the following identities
\bea
K^2 - K^{\mu \nu} K_{\mu \nu} &=& \hat{R} + 2 G_{\mu \nu} n^{\mu} n^{\nu}; \\
{\cal L}_{n} K_{\mu \nu} + K K_{\mu \nu} - 2 K_{\mu}^{\rho} K_{\rho \nu} &=& \hat{R}_{\mu \nu} - \gamma^{\rho}_{\mu} \gamma^{\sigma}_{\nu} R_{\rho \sigma} \nn,
\eea
with $G_{\mu \nu}$ the (bulk) Einstein tensor. One can simplify these expressions by fixing the gauge freedom such that $N = 1$ and $N^{\mu} =0$. In this gauge
\bea
ds^2 &=& dr^2 + \gamma_{ij} dx^i dx^j; \\
K_{ij} &=&  \frac{1}{2} \pa_r \gamma_{ij}. \nn
\eea
Moreover the Gauss-Codazzi relations which we will need can be written in terms of the trace adjusted stress energy tensor as
\bea
R_{rirj} &=& - \pa_r K_{ij} + K_{i}^{k} K_{kj}; \\
K^2 - K_{ij} K^{ij} &=& \hat{R} + \bar{T}_{rr} - \gamma^{ij} \bar{T}_{ij}; \nn \\
\pa_r K_{ij} - 2 K_{i}^{l} K_{lj} + K K_{ij} &=& \hat{R}_{ij} - \bar{T}_{ij}. \nn
\eea
Returning to \eqref{new-var} the terms in the variation which do not vanish given a Dirichlet condition on the metric, $\delta \gamma = 0$,
in this gauge take the form
\bea
\frac{2 \alpha}{\kappa^2} \int_{\partial {\cal M}} d^dx \sqrt{-\gamma} R_{rirj} \delta \gamma^{ij} &=&
\frac{2 \alpha}{\kappa^2} \int_{\partial {\cal M}} d^dx \sqrt{-\gamma} (-\pa_r K_{ij} + K_{i}^{k} K_{kj}) \pa_r \delta \gamma^{ij} \\
&=&
\frac{2 \alpha}{\kappa^2} \int_{\partial {\cal M}} d^dx \sqrt{-\gamma} (K K_{ij} - K_{i}^{k} K_{kj} - \hat{R}_{ij} + \bar{T}_{ij}) \pa_r \delta \gamma^{ij}, \nn
\eea
where in the last equality the bulk equation of motion in Gauss-Codazzi form has been used. Next we note that
\bea
\delta (K K^{ij} K_{ij}) &=& \frac{1}{2} K^{ij}K_{ij} \gamma^{kl} \pa_r \delta \gamma_{kl} + K K^{ij} \pa_r \delta \gamma_{ij} + \cdots \\
\delta (K^3)  &=& \frac{3}{2} K^2 \gamma^{ij} \partial_r \delta \gamma_{ij} + \cdots; \nn \\
\delta (\hat{R} K) &=& \frac{1}{2} \hat{R} \gamma^{ij} \partial_r \delta \gamma_{ij} + \cdots; \nn \\
\delta (K_{ij} K^{jk} K_{k}^{i}) &=& \frac{3}{2} K^{kj} K_{j}^{i} \partial_r \delta \gamma_{ij} + \cdots; \nn \\
\delta (\hat{R}^{ij} K_{ij}) &=& \frac{1}{2} \hat{R}^{ij} \pa_r \delta \gamma_{ij} + \cdots; \nn
\eea
where ellipses denote terms which do not depend on the normal derivative of the metric derivation. Then
\bea
&&\frac{2 \alpha}{\kappa^2} \int_{\partial {\cal M}} d^dx \sqrt{-\gamma} (K K_{ij} - K_{i}^{k} K_{kj} - \hat{R}_{ij} + \bar{T}_{ij}) \pa_r \delta \gamma^{ij} \\
&&
= \frac{2 \alpha}{\kappa^2} \int_{\partial {\cal M}} d^dx \sqrt{-\gamma} \delta (K K^{ij} K_{ij} - \frac{1}{3} K^3 + \hat{R} K - \frac{2}{3} K_{i}^{k} K_{kj} K^{ij} - 2 \hat{R}_{ij} K^{ij}) \nn \\
&& \qquad + \frac{2 \alpha}{\kappa^2} \int_{\partial {\cal M}} d^dx \sqrt{-\gamma} ( \gamma^{ij} (\bar{T}^{k}_{k} - \bar{T}_{rr}) + \bar{T}^{ij}) \pa_r \delta \gamma_{ij} + \cdots \nn
\eea
The terms in the second line are written in terms of quantities intrinsic to the boundary and define an analogue of the GHY term but the remaining terms left over in the last line cannot, in general, be expressed in terms of such quantities.

Suppose however that one works perturbatively in $\alpha$, evaluating the corrections
as a perturbative series on the leading order metric, so that
\be
g_{\mu \nu} = g^{(0)}_{\mu \nu}  + \alpha g^{(1)}_{\mu \nu} + \cdots
\ee
Working to order $\alpha$ in the action one needs to evaluate the terms involving $\bar{T}$ only to zeroth order in $\alpha$, i.e.
\be
( \gamma^{ij} (\bar{T}^{k}_{k} - \bar{T}_{rr}) + \bar{T}^{ij}) \rightarrow d \Lambda \gamma^{ij}
\ee
so that
\be
\int_{\partial {\cal M}} d^dx \sqrt{-\gamma} ( \gamma^{ij} (\bar{T}^{k}_{k} - \bar{T}_{rr}) + \bar{T}^{ij}) \pa_r \delta \gamma_{ij}
\rightarrow \int_{\partial {\cal M}} d^dx \sqrt{-\gamma} 2 d \Lambda \delta K + \cdots
\ee
That is, applying the field equations perturbatively, the problematic term can indeed be reexpressed in terms of quantities which are intrinsic to the boundary.

Putting the terms together, we see that working up to order $\alpha$ the boundary term needed to set up a well-defined Dirichlet variational problem at a finite radial boundary is
\bea
I_{GHY} &=& - \frac{1}{\kappa^2} \int_{\partial M} d^dx \sqrt{-\gamma} K  \nn  \\
&& - \frac{2 \alpha}{\kappa^2} \int_{\partial {\cal M}} d^dx \sqrt{-\gamma} \left (K K^{ij} K_{ij} - \frac{1}{3} K^3 + \hat{R} K  \right . \\
&& \qquad \qquad
\left . - \frac{2}{3} K_{i}^{k} K_{kj} K^{ij} - 2 \hat{R}_{ij} K^{ij} + 2 d \Lambda K \right ), \nn
\eea
where implicitly in the first line one needs the metric to order $\alpha$ whilst in the second line one needs the metric only to zeroth order in $\alpha$.

It would be interesting to extend this proof to show that the variational problem is well-defined to arbitrary order. To do this one would need to argue that
the problematic term
\be
\frac{2 \alpha}{\kappa^2} \int_{\partial {\cal M}} d^dx \sqrt{-\gamma} ( \gamma^{ij} (\bar{T}^{k}_{k} - \bar{T}_{rr}) + \bar{T}^{ij}) \pa_r \delta \gamma_{ij}
 \ee
can always be expressed as the variation of a term intrinsic to the boundary, when the bulk equations of motion are used iteratively. Such an all orders proof could be developed using similar inductive techniques to \cite{Compere:2011dx,Compere:2012mt}.

\section{Gravity models} \label{three}

In this section we will consider higher derivative corrections to Einstein gravity with a negative cosmological constant in four bulk dimensions. Before we describe the features of various models, let us comment on top down derivations of the effective action. One might think that it would be straightforward to work out the leading order corrections to the action from the reduction of ten or eleven dimensional actions, i.e. one could exploit our knowledge of the M theory action or the type II string actions. Here we point out that there are many subtleties in implementing such as strategy and our knowledge of these actions is not currently adequate to derive the corrections to AdS gravity actions in lower dimensions.

To illustrate this point let us consider the best understood top down possibility to obtain $AdS_4$, the reduction of M theory on a seven dimensional Sasaki-Einstein $SE_7$ to four dimensions. At the level of supergravity, it is always consistent to retain just the four-dimensional graviton in the lower dimensional theory, i.e. the eleven dimensional equations are solved by eleven-dimensional fields such that
\bea
ds_{11}^2 &=& ds^2_{4} (E_4) + ds_7^2(SE_7);  \label{eleven} \\
F_4 &=& \eta_{E_4}, \nn
\eea
where $E_4$ is any Einstein manifold with negative cosmological constant, $\eta_(E_4)$ is the volume form of this manifold and the metric reduction is diagonal over the $SE_7$. The effective four dimensional action is written only in terms of the metric on $E_4$, $g_{\mu \nu}$. Note however that not only the eleven-dimensional metric $g_{mn}$ but also the four form $F_4$ in eleven dimensions are non trivial, and the Riemann tensor of the $SE_7$ is also non trivial since the manifold has positive curvature.

Let us consider what this implies for the higher derivative corrections to the effective four-dimensional action. Since the four form is non-trivial at leading order, to compute the higher derivative corrections to the leading order solution, one would need to know higher derivative corrections to the eleven-dimensional action involving not just curvatures but also the four form.  Building on \cite{Howe:2003cy}, \cite{Cederwall:2004cg}, 
leading corrections involving the latter in eleven dimensions were worked out in
\cite{Hyakutake:2007sm}; they
have the structure
\bea
I &=& \int d^{11}x \sqrt{-g} a \left ( t_8 t_8 R^4 + \frac{1}{4!} \epsilon_{11}R^4 \right ) \\
&& + \int d^{11} x \sqrt{-g} b \left ( t_8 t_8 R^4 - \frac{1}{4!} \epsilon_{11}R^4 - \frac{1}{6} \epsilon_{11} t_8 A R^4  + [R^3 F^2] + [R^2 (DF)^2] \right ), \nn
\eea
where $a$ and $b$ are coefficients. It is known by comparison with IIA string calculations that
\be
b = \frac{1}{2 \kappa_{11}^2} \frac{l_{p}^6}{2^8 4!} \frac{\pi^2}{3},
\ee
with $2 \kappa_{11}^2 = (2 \pi)^8 l_p^9$. Here $\epsilon_{11}$ is the eleven-dimensional epsilon, $t_8$ consists of 4 Kronecker deltas and $t_8 t_8 R^4$ denotes a specific product of such such $t_8$ tensors and four Riemann tensors; the explicit expressions will not be needed here. $A$ is the three form of which $F$ is the four form field strength. The tensor structure of the terms denoted $[R^3 F^2]$ and $[R^2 (DF)^2]$ is also not important here; all we need is this schematic form, in which $D$ denotes the covariant derivative.

One might think that the knowledge of such terms would suffice to compute the leading corrections to the eleven-dimensional solution of interest, \eqref{eleven}, and that the these corrections could be rewritten in terms of a corrected equation for the four-dimensional metric $g_{\mu \nu}$, and hence in terms of a corrected four dimensional action. Apart from the complexity of the actual calculation, there would be a number of subtleties in actually carrying this out.

First of all, one cannot assume a priori that the higher order terms do not induce additional four-dimensional fields, as well as the metric, although it seems reasonable that in some cases they do not. For example, consider a four-dimensional massless scalar field $\phi$ which corresponds to a modulus of the dual conformal field theory. In principle, even though this field is constant at leading order, it could be sourced by a higher derivative correction, i.e. one could have an equation such as
\be
\Box \phi \sim R^n
\ee
where $R^n$ denotes schematically a scalar curvature invariant of order $n$. The latter must be zero when evaluated on AdS itself, as one does not expect the conformal invariance to be broken, but this argument could not exclude invariants of the Weyl tensor occurring.

Even if could argue that a four-dimensional action involving only $g_{\mu \nu}$ exists, there is a second obstacle in actually computing such an action To illustrate this point, consider just one of the tensor structures occurring in the $R^4$ invariant
\be
\frac{1}{l_p^3} (R_{mnpq} R^{mnpq})^2. \label{term1}
\ee
Evaluated on the lowest order metric this picks up contributions
\be
\frac{1}{l_p^3} \left ( (R_{\mu \nu \rho \sigma} R^{\mu \nu \rho \sigma})^2 +
(R_{\mu \nu \rho \sigma} R^{\mu \nu \rho \sigma})
(R_{abcd} R^{abcd}) + (R_{abcd} R^{abcd})^2 \right ),
\ee
where $R_{abcd}$ is the Riemann curvature of the Sasaki Einstein. For any given Sasaki Einstein this would then result in a term of the form
\be
I \sim \frac{V_{SE_7}}{l_p^3}  \int d^4 x \sqrt{-g}( (R_{\mu \nu \rho \sigma} R^{\mu \nu \rho \sigma})^2 +
b_2 (R_{\mu \nu \rho \sigma} R^{\mu \nu \rho \sigma}) + b_0), \label{mixing}
\ee
in the four-dimensional action with $V_{SE_7}$ the volume of the Sasaki-Einstein and $(b_0,b_2)$ computable (dimensionful) parameters. That is, a quartic invariant in eleven dimensions can lead to quadratic and constant terms in four dimensions, with the latter shifting the cosmological constant.

Since the Sasaki-Einstein has a curvature radius of the same order as the four dimensional manifold, none of these terms is subleading. Let $L$ be the scale of the curvature radius for both; then each of the three terms in the action is of order $L^3/l_p^3$, using the fact that Riemann squared is of order $1/L^4$. Note that the Einstein term in the action would be of order $L^9/l_p^9$. Terms arising from the reduction of higher order invariants in eleven dimensions would be subleading in a power series in $L/l_p$. 

Recall that in the case of $AdS_4 \times S^7$ the radius $L$ scales according to $L^6 \sim N l_p^6$ where $N$ is the rank of the dual gauge group. Therefore the Einstein term gives the well-known scaling of $N^{3/2}$ \cite{Henningson:1998gx} whilst the terms given above scale as $N^{1/2}$ and thus are suppressed by a factor of $1/N$ relative to the leading order terms. We will use this scaling later when discussing the spectrum. Similarly for the case of ABJM \cite{ABJM} where the eleven-dimensional geometry is $AdS_4 \times S^7/Z_k$ the curvature radius scales according to $L^6 \sim (k N) l_p^6$ where $N$ is the rank of the dual gauge group. Recalling that the volume of the compact space scales as $1/k$ this gives a scaling of $k^{1/2} N^{3/2}$ for the leading Einstein term. One can rewrite this as $k^2 \lambda^{3/2}$ where the 't Hooft coupling is $\lambda = N/k$, and this scaling was reproduced from the ABJM theory in \cite{Drukker:2010nc}. In the ABJM case the term give above would contribute at order $\lambda^{1/2}$, i.e. it differs by a factor of $1/(k^2 \Lambda)$ from the leading term. (Note that validity of the eleven-dimensional description requires $N \gg k^5$.)

 In conclusion, identifying the leading order corrections in four dimensions is very subtle.  The leading order correction in four dimensions indeed derives from the leading order correction in eleven dimensions, but terms involving higher curvature invariants in four dimensions can actually contribute at the same order! Similarly terms in the higher dimensional action involving $R^3 F^2$ and so on can give rise to corrections in four dimensions involving $R^3$.

Note that the term picked out above \eqref{term1} shifts the cosmological constant and is non zero even when evaluated on AdS itself. This would mean, in particular, that it would be expected to adjust the value of the free energy for the dual theory (at zero temperature) evaluated on an $S^3$. If one can argue that there is no such renormalisation, then the four-dimensional contributions from such a term must cancel those arising from the reduction of other eleven-dimensional terms. A series of corrections expressed in terms of the Weyl tensor, which vanishes on a maximally symmetric space, would not induce such a change in the free energy. 

From the string theory perspective one might think that one should in any case start from a higher dimensional with curvature corrections involving only the Weyl tensor, since corrections involving Ricci and Ricci scalar can always be absorbed into field redefinitions. Here we will look nonetheless look at terms such as \eqref{term1} as well as Weyl terms. Firstly it is is interesting to look at the effects of different curvature invariant structures but secondly the usual field redefinition argument refers to the bulk field equations but does not take into account boundary conditions and onshell thermodynamic quantities. We will see below that it is possible to have terms which do not contribute to the field equations perturbed around a given leading order solutions but which nonetheless change the action and change the spectrum. In particular, curvature squared corrections in four dimensions do not change the metric, so in the past they would have been viewed as trivial, but here we show that they still introduce additional (non-unitary) operators into the dual CFT spectrum.

In what follows, we will pursue a bottom up perspective, in which we consider case by case the effects of various higher derivative terms in four dimensions.  In other words, we discuss the effects of adding particular scalar curvature invariants to the four dimensional action. We will then return to the issue of which scalar invariants are expected to arise in top down models.

\subsection{Curvature squared corrections}

Motivated by requirements of renormalizability of gravity, curvature-squared modifications
to Einstein's theory were first discussed in \cite{Stelle1, Stelle2} and they have been extensively explored in the literature. The most general action involves curvature squared terms can be written as
\be
I = \frac{1}{2 \kappa^2} \int d^4 x \  \sqrt{-g} (R - 2\L + \a R^{\m\n}R_{\m\n} + \b R^2 + \g R^{\m\n\r\s}R_{\m\n\r\s})
\ee
However, it is well known that the Gauss-Bonnet invariant,
\be
E_4 = R^{\m\n\r\s}R_{\m\n\r\s} - 4R^{\m\n}R_{\m\n} + R^2
\ee
does not contribute to the equations of motion in four dimensions but yields only a surface term.
Hence for analysing the field equations we can eliminate the Riemann squared term in the action above, making the most general action we need to consider, modulo the $E_4$ term, simply
\be \label{GaussBonnetAction}
I = \frac{1}{2 \kappa^2} \int d^4 x \  \sqrt{-g} (R - 2\L + \a R^{\m\n}R_{\m\n} + \b R^2),
\ee
(where implicitly the coefficients $(\alpha, \beta)$ have been shifted relative to the above.)
The equations of motion following from this action are
\be \label{GaussBonnetEom}
{\cal G}_{\m\n} + E_{\m\n} = 0
\ee
where
\be \label{gmn}
{\cal G}_{\m\n} = R_{\m\n} - \frac{1}{2} R g_{\m\n} + \L g_{\m\n}
\ee
and
\bea \label{emn}
E_{\m\n} &=& 2\a (R_{\m\r}R_\n^\r - \frac{1}{4}R^{\r\s}R_{\r\s}g_{\m\n} ) + 2\b R(R_{\m\n} - \frac{1}{4}R g_{\m\n}) \\
                && + \a (\Box R_{\m\n} + \frac{1}{2} \Box R g_{\m\n} - 2\nabla_\r \nabla_{(\m} R_{\n )}^\r ) + 2\b (g_{\m\n} \Box R - \nabla_\m \nabla_\n R) \nn
\eea
By analyzing the above equations of motion it immediately follows that
all solutions of the $\a = \b = 0$ theory are also solutions of the full theory as
$E_{\m\n}$ is zero for any Einstein spacetime. In particular,
AdS-Schwarzschild black holes
\be \label{ads-bh}
ds^2 = - dt^2 ( \ep - \frac{m}{r} + \frac{|\Lambda|}{3} r^2 )
+ \frac{dr^2}{( - \frac{m}{r} + \frac{|\Lambda|}{3} r^2 + \ep)} + r^2 d\Omega_2^2(k)
\ee
are solutions of the higher curvature theory. Here $k=0$ and $\ep=0$ corresponds to the case in which the horizon is flat, with
$k=1$ and $\ep =1$ corresponding to the case in which the horizon is a two-sphere. Note however that
the thermodynamic properties are modified in the deformed theory and depend explicitly on the deformation parameters.

It is straightforward to derive the thermodynamic properties in the deformed theory, exploiting the fact that the metric remains
Einstein. (For earlier discussions of thermodynamics in bulk dimensions higher than four see  \cite{Nojiri:1999nd}.)
The free energy of the black holes can be obtained by considering the onshell value of the action.
In order for the variational problem to be well-defined, the action must be supplemented
by boundary terms. For the Einstein part of the action the appropriate boundary term is the Gibbons-Hawking-York term discussed earlier
\be
I_{GHY} = - \frac{1}{\kappa^2} \int d^3x K \sqrt{- \gamma}.
\ee
where $K$ denotes the second fundamental form and $\gamma$ is the boundary metric. This term is not however sufficient to ensure a well-defined variational problem: in varying the bulk action the following boundary terms arise, analogously to those given in \eqref{new-var}
\be
\frac{1}{\kappa^2} \int_{\partial {\cal M}} d\Sigma^{\mu} \left ( 2 \alpha R_{\nu \sigma} + 2 \beta R g_{\nu \sigma} \nabla_{\mu} \delta g^{\sigma \nu}  + \cdots \right )
\ee
where we again suppress terms which vanish for the boundary condition $\delta g = 0$. Using the equation of motion $R_{\mu \nu} = \Lambda g_{\mu \nu}$ we note that the variational problem will be well posed if we add the following boundary terms
\be
I = - \frac{1}{\kappa^2} \int d^3x \sqrt{-\gamma} K (2 \alpha  \Lambda + 8 \beta \Lambda). \label{bn1}
\ee
In this case the fact that the solution remains Einstein implies that this term is sufficient to evaluate the onshell action, to all perturbative orders in $\alpha$ and $\beta$. (It does not however suffice for discussing fluctuations around the Einstein solution, as we will discuss in section four.)

Evaluating the complete onshell action gives
\bea
I &=& \frac{1}{2 \kappa^2} 2 \Lambda (1 + 2 \alpha \Lambda + 8 \beta \Lambda) \int d^4 x \sqrt{-g} \\
&& - \frac{1}{\kappa^2} (1 + 2 \alpha \Lambda + 8 \beta \Lambda) \int d^3 x \sqrt{-\gamma} K. \nn
\eea
Relative to the case of $\alpha = \beta = 0$, there is just an overall prefactor, which means that we can immediately
read off from \cite{Henningson:1998gx,Balasubramanian:1999re,de Haro:2000xn}
the required holographic counterterms as
\be
I_{ct} = - \frac{1}{2\kappa^2} (1 + 2 \alpha \Lambda + 8 \beta \Lambda) \int d^3 x \sqrt{-\gamma} \left (
\frac{4}{l}  + l {R} (\gamma) \right ),
\ee
where $l^2 = |3/\Lambda|$.
The asymptotic expansion of the metric ${g}$ is \cite{Henningson:1998gx,de Haro:2000xn}
\bea
ds^2 &=& l^2 \left( \frac{d \rho^2}{\rho^2} + \frac{1}{\rho^2} {g}_{ij}(x,\rho) dx^i dx^j \right ); \label{fg0} \\
{g}_{ij}(x,\rho) &=& {g}_{(0) ij} (x) + \rho^2 {g}_{(2) ij} (x) + \rho^3 {g}_{(3) ij} + \cdots, \nn \\
{g}_{(2) ij} &=& - {R}_{ij} ({g}_{(0)}) + \frac{{R}({g}_{(0)})}{4} {g}_{(0)ij}, \nn
\eea
with ${g}_{(3)}$ being traceless and divergenceless but otherwise undetermined by the asymptotic analysis.
The renormalized stress energy tensor obtained by varying the action with respect to ${g}_{(0)}$ is then
shifted by an overall prefactor relative to \cite{de Haro:2000xn}
\be
\langle T_{ij} \rangle = \frac{3}{2 \kappa^2}  (1 + 2 \alpha \Lambda + 8 \beta \Lambda) {g}_{(3) ij}.
\ee
We can now immediately evaluate thermodynamic quantities for the black hole solutions \eqref{ads-bh}; the free
energies and masses are clearly shifted relative to those in Einstein gravity by a proportionality factor:
\bea
- \beta_T F & \equiv & I^E_{{\rm onshell}} = \frac{\beta_T  V_{xy}}{2 \kappa^2} (1 + 2 \alpha \Lambda + 8 \beta \Lambda) m \\
M & \equiv & \int d^2x \sqrt{- \gamma} \langle T_{00} \rangle
= \frac{V_{xy}}{\kappa^2} (1 + 2 \alpha \Lambda + 8 \beta \Lambda) m
\eea
with $\beta_T$ the inverse temperature (not to be confused with the coupling constant $\beta$) and the temperature being
\be
T = \frac{1}{4 \pi} \left ( \frac{2 | \Lambda |}{3} r_h + \frac{m}{r_h^2} \right ),
\ee
and $r_h$ is the horizon position. $I^E$ denotes the Euclidean action, which in this static case is straightforwardly computed by analytic continuation of the time. Note that under such a continuation
$i I \rightarrow - I^E$. One can also work out the black hole entropy using Wald's method \cite{Wald:1993nt}. Define
\be
{\cal Q}^{\m\n} = -2 {\cal L}^{\m\n\r\s} \nabla_\r \  l_\s + \cdots
\ee
where ${\cal Q}^{\m\n}$ is antisymmetric and the terms denoted by ellipses vanish for stationary horizons.
Here
\be
{\cal L}^{\m\n\r\s} \equiv \frac{\delta L}{\delta R_{\m\n\r\s}}.
\ee
For a stationary horizon the black hole entropy is then given by
\be \label{wald}
S = \frac{1}{T} \int_{{\cal H}} {\cal Q}^{\m\n}\ d\Sigma_{\m\n},
\ee
with $T$ being the horizon temperature, ${\cal H}$ denoting the horizon and $l^{\k}$ being the horizon normal.

For Einstein gravity
\be
L = \frac{1}{2 \kappa^2} \sqrt{-g} \left( R - 2\L \right)
\ee
where $\kappa^2 = 8 \pi G$ and $G$ is the Newton constant. Hence
\be
\frac{\delta L}{\delta R_{\m\n\r\s}} = \frac{\sqrt{-g}}{4 \kappa^2} \left( g^{\m\r} g^{\n\s} - g^{\m\s} g^{\n\r} \right)
\ee
And so
\be
{\cal Q}^{\m\n} = -\frac{\sqrt{-g}}{2 \kappa^2} \left( \nabla^\m l^\n - \nabla^\n l^\m \right)
\ee
and
\be
S = \frac{1}{\kappa^2 T} \int_{{\cal H}} \sqrt{-g} \left( \nabla^\n l^\m \ d\Sigma_{\m\n} \right) \equiv \frac{A_h}{4 G}
\ee
with $A_h$ the horizon area, using $l_{\nu} \nabla^{\nu} l^{\mu} = \kappa_h l^{\mu}$ where $\kappa_h = 2 \pi T$ is the surface gravity
of the horizon.

For the curvature squared corrections, using
\bea
\frac{\delta R^2}{\delta R_{\mu \nu \rho \sigma}} &=& R (g^{\mu \rho} g^{\sigma \nu} - g^{\mu \sigma} g^{\nu \rho}); \\
\frac{\delta (R^{\tau \eta} R^{\tau \eta})}{\delta R^{\mu \nu \rho \sigma}} &=& ( R^{\mu \rho} g^{\sigma \nu} -
R^{\mu \sigma} g^{\nu \rho}), \nn
\eea
the Wald entropy becomes
\be
S = (1 + 2 \alpha \Lambda + 8 \beta \Lambda) \frac{A_h}{4 G}.
\ee
Putting these results together one finds that the thermodynamic relations
\be
F = M - T S; \qquad dM =T dS,
\ee
are indeed satisfied.

To summarise, the metric is uncorrected but the thermodynamic properties of the black holes are adjusted: the entropy, the temperature, the mass and the free energy are all changed, albeit by just an overall factor.
 It is also interesting to note that the action evaluated on $AdS_4$ with $S^3$ boundary is also changed. In the latter case the relevant bulk metric is
\be
ds^2 = \frac{3}{|\Lambda|} \left ( \d\r^2 + \sinh^2\r \ \d \O_3^2 \right ),
\ee
where $0<\r<\infty$. Using the renormalised action given above, one can compute the onshell Euclidean action to be
\be
I^E_{\rm onshell} =  \frac{12 \p^2}{ |\Lambda| \k^2}(1+2\L\a+8\L\b). \label{cc2}
\ee
This is therefore corrected by the curvature squared terms except when $\alpha = - 4 \beta$, which corresponds to the case in which the correction is Riemann squared minus $E_4$. Given a holographic dual in which one can compute the free energy on $S^3$ by localisation techniques, the answer will give a criterion restricting the terms which can arise in the effective four-dimensional action. In particular, the case of ABJM theory, for which the exact expression for the planar free energy was obtained in \cite{Drukker:2010nc}, will be explored in detail elsewhere. 

\bigskip

At this point it would seem as if the addition of such terms to the action is rather trivial because the thermodynamic quantities are shifted by an overall factor, which could be reabsorbed into the cosmological constant. However, we will discuss in section \ref{spectrum}, these terms are highly non-trivial when one looks at the spectrum of the theory.
To find the spectrum of the dual CFT
linearize the above field equations about the background solution $AdS_4$.
As we discuss in section \ref{spectrum}, the bulk theory is found to describe a massless spin-2 graviton, a massive scalar and
a massive spin-2 field. By tuning the coefficients
so that $\a = -3\b$ one may eliminate the massive scalar mode. One can also tune
the remaining coefficient $\b$ to the so-called critical value \cite{Pope1}
\be
\beta = - \frac{1}{2 \Lambda} \label{crt}
\ee
where the massive spin two mode becomes logarithmic \cite{Johansson:2012fs}. Noting that the AdS-Schwarzschild
mass when $\a = - 3 \b$ behaves as
\be
M = \frac{m}{\kappa^2} (1 + 2 \beta \Lambda)
\ee
we see that in the critical theory the black hole solution has zero mass. One can show
furthermore that the Wald entropy vanishes at the critical point.

However we should emphasize that
the critical value \eqref{crt} can clearly never be achieved when the Planck length is small compared to
the curvature radius of AdS. If one is viewing
the higher curvature corrections as arising from a top down string model then $\beta \Lambda$ is necessarily much smaller than one.
The critical theory does not therefore provide a good model for corrections to macroscopic $AdS_4$ black holes.

Let us now make a connection to conformal gravity.
With the first parameter choice of $\a = -3 \b$ we may rewrite the higher curvature term in terms of the Weyl tensor:
\be
-\frac{1}{3} \a (R^2 - 3R^{\m\n}R_{\m\n}) = \frac{1}{2} \a (C^{\m\n\r\s}C_{\m\n\r\s} - E_4)
\ee
Hence the Lagrangian is equivalent to
\be
I = \frac{1}{2 \kappa^2} \int d^4 x \  \sqrt{-g} \left(R - 2\L + \frac{1}{2} \a (C^{\m\n\r\s}C_{\m\n\r\s} - E_4) \right).
\ee
By taking the limit of $\alpha \rightarrow \infty$ one recovers Weyl gravity, see related discussions in \cite{Maldacena}, but again this would not be reached as a small correction from an Einstein solution.

The variation of the Weyl squared term with respect to the metric is linear in the Weyl tensor. This means that this correction vanishes identically when evaluated on $AdS_4$ since it has vanishing Weyl tensor. Therefore we can deduce from \eqref{cc2} that the onshell Euclidean action evaluated on $AdS_4$ with $S^3$ boundary for
\be
I^E = - \frac{1}{2 \kappa^2} \int d^4 x \sqrt{g} \left ( R - 2 \Lambda - \frac{1}{2} \alpha E_4 \right)
\ee
is
\be
I^E_{\rm onshell} = \frac{12 \pi^2}{|\Lambda| \kappa^2} (1 - \frac{2 \alpha \Lambda}{3}).
\ee
In other words, the topological invariant does of course contribute to the action even though it does not affect the field equations. The renormalised $E_4$ term captures the Euler invariant of the manifold with $S^3$ conformal boundary. Tuning to the critical value \eqref{crt} this action is zero.

\subsection{$f(R)$ Gravity} \label{f(R)Gravity}

In our exploration of corrected black hole solutions we will now move on to consider an
$f(R)$ theory. The $f(R)$ theory is obtained when we add a generic polynomial in the Ricci scalar $R$ to the usual Einstein action,
\be
I = \frac{1}{2 \kappa^2} \int d^4 x   \sqrt{-g} \left ( R - 2 \L + f(R) \right )
\ee
where
\be
f(R) = \sum_{n \ge 2} \a_{n} R^n,
\ee
with arbitrary coefficients $\a_n$. There us considerable interest in $f(R)$ theories in the context of phenomenology and cosmology, see the reviews of \cite{Sotiriou:2008rp,DeFelice:2010aj,Nojiri:2010wj}, even though such corrections are not well motivated from top down considerations, since they can be removed by field redefinitions.

Indeed it is well known that such a correction will not change the leading order black hole solution non-trivially,
although it will change its thermodynamic properties. This follows from the equations of motion
\be \label{fe2}
{\cal G}_{\m\n} + F_{\m\n} = 0
\ee
with ${\cal G}_{\m \n}$ defined in \eqref{gmn} and
\bea
F_{\m\n} &=& \sum_{n \ge 2} \a_n n  R^{n-1} (R_{\m\n} - \frac{1}{2 n}R g_{\m\n}) \\
&& +  \sum_{n \ge 2} \a_n n (g_{\m \n} \Box R^{n-1} - \nabla_{\mu} \nabla_{\nu} R^{n-1}). \nn
\eea
Consider an Einstein solution $g_{\m \n}$ which satisfies
\be
R_{ \mu \nu} = \lambda g_{\m \nu}; \qquad
R_{ \mu \nu} = \frac{1}{4} R g_{\mu \nu},
\ee
Evaluated on such a solution
both the second line together with the $n=2$ term in the first line of $F_{\m \n}$ vanish and
\be
F_{\m\n} = \sum_{n > 2} \a_n (n-2) (4 \lambda)^{n-1} \lambda g_{\mu \nu} \equiv - \delta \Lambda g_{\m \n}.
\ee
The field equations \eqref{fe2} are then satisfied provided that
\be
\Lambda = \lambda + \sum_{n > 2} \a_n (2-n) (4 \lambda)^{n-1} \lambda;
\ee
i.e. the higher derivative term acts to shift the effective cosmological constant. Treating the $f(R)$ term as a small perturbation
around the leading order solution by setting all coefficients $\alpha_n \ll 1$, we may express
\be
\lambda \approx \Lambda (1 - \a_3 (4 \Lambda)^2 - 2 \a_4 (4 \Lambda)^3 + \cdots),
\ee
where in the non-linear terms we use the leading order behavior $\lambda \sim \Lambda$.
Therefore any Einstein solution remains an Einstein solution in the corrected theory, but with a shifted cosmological constant.

One should again note that the corrected theory does admit non-Einstein solutions, but any solution which reduces to
an Einstein solution in the leading order theory remains Einstein in the corrected theory. In other words, when one treats the higher order
terms perturbatively one discards solutions which do not reduce to Einstein solutions on setting $\a_{n}$ to zero. The higher derivative terms with recur, however, when one discusses the spectrum as one will obtain new propagating modes.

\bigskip

As a warm up exercise for the non-trivial corrections discussed in the following sections it is useful to
derive the corrections to static solutions as follows, using a similar method to that of \cite{Deser} and also
 \cite{deHaro:2003zd}. Let the metric be parameterized as
\be \label{DeserMetric}
ds^2 = - a(r) b^2(r) dt^2 + \frac{dr^2}{a(r)} + r^2 (dx^2 + dy^2),
\ee
where we now focus on the case of flat horizons. Substituting this metric ansatz, the action reduces to
\be \label{bulk}
I = \frac{1}{2 \kappa^2} \beta_T V_{xy} \hat{I}
\ee
with $\beta_T$ the periodicity in time; $V_{xy}$ the regulated volume of the $(x-y)$ plane and
\bea
\hat{I} &=& - \int dr   \left[ 2rb^{\prime}( - \frac{\Lambda}{3} r^2 - a) +
\sum_n (-1)^n \a_{n} \frac{A^n}{(br^2)^{n-1}}  \right] \label{Ect} \\
&& \qquad - \left [ \frac{2}{3} \Lambda b r^3 + 2 a b^{\prime} r^2 + a^{\prime} b r^2 + 2 a b r \right ]_{r_h}^{\infty} \nn
\eea
where we have used the fact that the Ricci scalar is given by
\be
R = - \frac{A(r)}{ b(r) r^2}
\ee
with $A(r)$ given by
\be
A(r) \equiv 3a^{\prime}b^{\prime}r^2 + 2ab^{\prime\prime}r^2 +a^{\prime\prime}br^2 + 4a^{\prime}br +4ab^{\prime}r + 2ab.
\ee
The second line in \eqref{Ect} arises from partial integrations.
Varying the bulk term in the action we find the following equations of motion for $a$ and $b$:
\be
0 = 2rb' - \sum_n l^{n-1} (-1)^n \a_n \frac{\delta}{\delta a} \left( \frac{A^n}{(b r^2)^{n-1}} \right )
\ee
where
\bea
\frac{\delta}{\delta a} \left( \frac{A^n}{(b r^2)^{n-1}} \right ) &=& \frac{nA^{n-1}}{(br^2)^{n-1}}(2b''r^2+4b'r+2b) - \frac{d}{dr} \left( \frac{nA^{n-1}}{(br^2)^{n-1}}(3r^2b'+4b) \right) \nn \\
&&+ \frac{d^2}{dr^2} \left( \frac{nA^{n-1}}{(br^2)^{n-1}}(br^2) \right)
\eea
The other equation of motion is
\be
0 = - 2 \Lambda r^2 - 2a - 2a'r  - \sum_{n} l^{n-1} (-1)^n \a_n \frac{\delta}{\delta b} \left( \frac{A^n}{(b r^2)^{n-1}} \right )
\ee
where
\bea
 \frac{\delta}{\delta b} \left( \frac{A^n}{(b r^2)^{n-1}} \right )
&=& \frac{(1-n)A^n}{r^{2n-2}b^n} + \frac{nA^{n-1}}{(br^2)^{n-1}}(a''r^2+4a'+2a) \\
&& - \frac{d}{dr} \left( \frac{nA^{n-1}}{(br^2)^{n-1}}(3r^2a'+4ar) \right)
+ \frac{d^2}{dr^2} \left( \frac{nA^{n-1}}{(br^2)^{n-1}}(2ar^2) \right). \nn
\eea
Solving these equations of motion to linear order in the coupling constants $\a_i$ we expand as:
\bea
a(r) &=& a_{(0)}(r) + \sum_n \a_n a_{(n)}(r); \\
b(r) &=& b_{(0)}(r) + \sum_n \a_n b_{(n)}(r). \nn
\eea
To leading order, namely all $\a_n = 0$, the equations are solved by
\be
b_{(0)} (r) = 1; \qquad
a_{(0)}(r) = - \frac{1}{3} \Lambda r^2 - \frac{m}{r}
\ee
To linear order in the perturbations, the general solution to the equations of motion is
\bea
b_{(n)}(r) &=& 0; \\
a_{(n)}(r) &=& (-)^n (n-2) (4 \Lambda)^{n-1} \Lambda r^2 - \frac{m_{n}}{r}. \nn
\eea
The latter renormalizes the cosmological constant and in addition allows for a shift in the integration constant which
parameterizes the black hole mass: the corrected metric is
\be
ds^2 = \frac{dr^2}{ (- \frac{1}{3} \lambda r^2 - \frac{m}{r} - \sum_n \frac{\alpha_n m_{n}}{r})}
- dt^2  (- \frac{1}{3} \lambda r^2 - \frac{m}{r} - \sum_n \frac{\alpha_n m_{n}}{r}) + r^2 dx \cdot dx. \label{bh-met}
\ee

\subsubsection{Black hole thermodynamics in $f(R)$ theory}

For the $f(R)$ term the analysis is of the variational problem is subtle: varying the bulk term gives rise to a boundary variation
\be
\delta I = \frac{1}{\kappa^2} \int d^3x f^{\prime} (R) \delta (K \sqrt{-\gamma}),
\ee
where $f^{\prime} (R) = \pa_R f(R)$. The appropriate boundary term
for a four-dimensional $f(R)$ theory was argued by Hawking and Luttrell \cite{Hawking:1984ph}
to be
\be
I_{HL} = - \frac{1}{\kappa^2} \int d^3 x f^{\prime} (R) K \sqrt{-\gamma}
\ee
However, in general this is not satisfactory since $R$ is not intrinsic to the boundary, i.e. it is the scalar curvature of the bulk metric, rather than the boundary metric. In the case at hand however one can use the fact that the onshell Ricci scalar is constant to write this term in terms of quantities manifestly intrinsic to the boundary.
Putting all terms together the complete action is
\bea
I &=& \frac{1}{2 \kappa^2} \int d^{4}x \sqrt{-g} \left ( R - 2 \Lambda + f(R) \right ) \\
&& - \frac{1}{\kappa^2} \int d^3 x \sqrt{-\gamma} K \left (1 + f^{\prime}(R) \right ). \nn
\eea
To evaluate the free energy one needs to holographically renormalize this action. To linear order in
the couplings for the higher derivative terms, however, one can immediately carry out this procedure using the known results
for asymptotically locally AdS Einstein manifolds. Recall that the deformed solution is Einstein, with a different
cosmological constant, the metric can always be expressed as
\be
ds^2 = \tilde{l}^2 d\bar{s}^2
\ee
with
\be
\tilde{l}^2 = \frac{3}{ |\lambda|}; \qquad
\bar{R}_{\mu \nu} = - 3 \bar{g}_{\mu \nu}
\ee
and $\lambda < 0$.
The asymptotic expansion of the metric $\bar{g}$ is known
\bea
ds^2 &=& \frac{d \rho^2}{\rho^2} + \frac{1}{\rho^2} \bar{g}_{ij}(x,\rho) dx^i dx^j; \label{fg1} \\
\bar{g}_{ij}(x,\rho) &=& \bar{g}_{(0) ij} (x) + \rho^2 \bar{g}_{(2) ij} (x) + \rho^3 \bar{g}_{(3) ij} + \cdots, \nn \\
\bar{g}_{(2) ij} &=& - \bar{R}_{ij} (\bar{g}_{(0)}) + \frac{\bar{R}(\bar{g}_{(0)})}{4} \bar{g}_{(0)ij}, \nn
\eea
with $\bar{g}_{(3)}$ being traceless and divergencless but otherwise undetermined by the asymptotic analysis. Given this form
for the asymptotic expansion one can now compute the regulated action and hence the counterterms. In doing so one can
use the fact that to linear order in the new couplings
\be
f(R) \rightarrow f(R)|_{R = 4 \Lambda}.
\ee
Setting $\Lambda = -3$ so that the metric to leading order is normalized to unit curvature radius,
the required counterterms are then
\bea
I_{ct} &=& - \frac{1}{2\kappa^2} (1 + \sum_{n} 2 \alpha_n (-12)^{n-1}) \int d^3 x \sqrt{- \bar{\gamma}} \left (
4 + \bar{R} (\bar{\gamma}) \right ); \\
&=& \frac{1}{2 \kappa^2}
(1 + \sum_{n} (\frac{3n}{2} -1) \alpha_n (-12)^{n-1}) \int d^3 x \sqrt{- \bar{\gamma}} \left (
4 \right ) \nn \\
&& +  \frac{1}{2 \kappa^2}
(1 + \sum_{n} (\frac{3n}{2} -1) \alpha_n (-12)^{n-1}) \int d^3 x \sqrt{- \bar{\gamma}}  (1 + \sum_{n} (n-2) \alpha_n (-12)^{n-1}) \bar{R} (\bar{\gamma}) , \nn
\eea
and the renormalized stress energy tensor obtained by varying the action with respect to $\bar{g}_{(0)}$ is
\be
\langle T_{ij} \rangle = \frac{3}{2 \kappa^2}  (1 + \sum_{n} 2 \alpha_n (-12)^{n-1}) \bar{g}_{(3) ij}.
\ee
One can then compute the mass of the black hole in \eqref{bh-met} as
\be
M = \int d^2 x \langle T_{00} \rangle = \frac{V_{xy}}{2 \kappa^2} (1 + \sum_{n} 2 \alpha_n (-12)^{n-1})
(m + \sum_n \alpha_n m_n),
\ee
and evaluating the onshell action gives
\be
- \beta_T F = I^E \equiv  = \beta_T \frac{V_{xy}}{2 \kappa^2} (1 + \sum_{n} 2 \alpha_n (-12)^{n-1})  (m + \sum_n \alpha_n m_n)
\ee
with $F$ the free energy and the black hole temperature being $1/\beta_T$.

In the $f(R)$ theory using the fact that
\be
\frac{\delta R^n}{\delta R_{\m\n\r\s}} = \frac{1}{2} n R^{n-1} (g^{\m \r} g^{\n \s} - g^{\m \s} g^{\n \r}),
\ee
the Wald entropy \eqref{wald} is given by
\be
S = (1 + \sum_n \alpha_n (4 \Lambda)^{n-1}) \frac{2 \pi A_h}{\kappa^2}.
\ee
Evaluating this one obtains
\be
S = \frac{2 \pi V_{xy}}{\kappa^2} (1 + \sum_n 2 \alpha_n (-12)^{n-1}) (m + \sum_n \alpha_n m_n)^{2/3}.
\ee
Finally the temperature of the black hole is given by
\be
T = \frac{3}{4 \pi} (m + \sum_n \alpha_n m_n)^{1/3}.
\ee
Putting these results together one sees that the relation $F = M - TS$ is satisfied together with the first
law $dM = T dS$. Moreover, it is clear that by choosing the integration constants $m_n$ such that
\be
m_n = - 2 (-12)^{n-1} m
\ee
the black hole in the $f(R)$ theory has unchanged thermodynamic properties to leading order in the coupling constants
$\alpha_n$.

\subsection{Einstein + $C^3$} \label{WeylCubed}

We now move on to consider the addition to the action of curvature invariants of degree three or higher. At this point it is useful to look at classifications of scalar curvature invariants in our dimensions. One such set of invariants are the Carminati-McLenaghan invariants, \cite{Carminati}. At degree three the possible invariants include both those built of lower degree invariants, for example $R C^{\m\n\r\s} C_{\m\n\r\s}$, and invariants built by contracting three tensors with each other. At degree three the latter gives the new invariants
\bea
&& S^{\mu \rho} S_{\rho \nu} S^{\nu}_{\mu} \qquad
C^{\mu \nu}_{\ \ \rho \sigma} C^{\rho \sigma}_{\ \ \tau \eta} C^{\tau \eta}_{\ \ \mu \nu} \qquad
\ast C^{\mu \nu}_{\ \ \rho \sigma} C^{\rho \sigma}_{\ \ \tau \eta} C^{\tau \eta}_{\ \ \mu \nu} \\
&& S^{\mu \nu} S_{\rho \sigma} C_{\mu \rho \nu \sigma} \qquad
S^{\mu \nu} S_{\rho \sigma} \ast C_{\mu \rho \nu \sigma} \nn
\eea
where $S_{\mu \nu}$ is the traceless Ricci tensor, $C_{\mu \nu \rho \sigma}$ is the Weyl tensor and
$\ast C_{\mu \nu \rho \sigma}$  denotes the dual of the Weyl tensor. Note that these comprise an over complete set of invariants for a planar static spacetime. Curvature invariants built from the Ricci scalar or Ricci tensor will behave qualitatively similarly to those at quadratic order, leaving the metric unchanged but shifting the action. Therefore in this section we will focus on the effect of the
cube of the Weyl tensor on planar black hole solutions, which is qualitatively different.

The action we consider is therefore
\be
I = \frac{1}{2 \kappa^2} \int d^4 x \  \sqrt{-g} \left ( R - 2\L + \a C_{\m\n}^{\ \ \r\s}C_{\r\s}^{\ \ \eta\l}C_{\eta\l}^{\ \ \m\n}  \right ) \label{bulk-weyl}
\ee
where $C_{\m\n\r\s}$ is the Weyl tensor.

Since the general field equations are somewhat complicated in this case, the easiest way to obtain the corrections to the planar black holes is as follows.
Evaluated on a static ansatz \eqref{DeserMetric} the action reduces to and effective one dimensional action
\bea
\hat{I} &=& \int_{r_H}^\infty dr\  \left[ 2rb^{\prime}(a+\frac{\L}{3}r^2) - \a \frac{B^3}{18r^4b^2} \right] \nn \\
&& \qquad - \left [ \frac{2}{3} \Lambda b r^3 + 2 a b^{\prime} r^2 + a^{\prime} b r^2 + 2 a b r \right ]_{r_h}^{\infty}
\eea
where
\be
B \equiv 2ab - 2rba' - 2rab' + 3r^2a'b' + r^2ba'' + 2r^2ab''
\ee
Varying this action we get the equations of motion for $a$ and $b$ and these are solved perturbatively in $\a$ as in Sections \ref{f(R)Gravity} and \ref{RiemannFour}. There is in this case a non-trivial correction to the planar black hole solution:
\be
a_{(1)}(r) = \frac{a_1}{r} - \frac{8m^2\L}{r^4} - \frac{16m^3}{r^7}
\ee
and
\be
b_{(1)}(r) = b_1 -6\frac{m^2}{r^6}.
\ee
Here $a_1$ and $b_1$ are arbitrary integration constants. The former acts as a redefinition of the mass parameter $m$ at order $\alpha$ and the latter changes the norm of the time Killing vector at infinity at order $\alpha$. We will discuss the interpretation of these integration constants further below but setting them to zero we obtain
\bea
a(r) &=& a_{(0)}(r) + \a a_{(1)}(r)\nn \\
       &=& -\frac{\L}{3}r^2 -\frac{m}{r} + \a \left( - \frac{8m^2\L}{r^4} - \frac{16m^3}{r^7} \right),  \label{aWeyl} \\
       && \nn \\
b(r) &=& b_{(0)}(r) + \a b_{(1)}(r)\nn \\
       &=& 1 + \a \left( -6\frac{m^2}{r^6} \right). \nn
\eea
Note that the AdS solution itself is uncorrected, as one would expect: the contribution to the field equations from the variation of the Weyl cubed term is given below in \eqref{weyl1} and evaluated on a solution with vanishing Weyl tensor it is zero.

\subsubsection{Thermodynamics of corrected black hole solutions}

Let us now work out the thermodynamics of the corrected black hole solution.
The horizon is given by $r_H$ such that $a(r_H)=0$. Since $a(r) = a_{(0)}(r) + \a a_{(1)}(r)$, we find $r_H$ also to order $\alpha$.
Let
\be
r_H = r_{H(0)} + \a r_{H(1)}
\ee
where
\bea
r_{H(0)}^3 &=& -\frac{3m}{\L}; \\
r_{H(1)} &=& r_{H(0)} \left( -\frac{2^3}{3^3}\L^2 \right). \nn
\eea
The temperature of this black hole solution is given by:
\bea
T &=& \frac{a'(r_H)b(r_H)}{4\pi}  \\
   &=& \frac{|\L|}{4\pi} r_{H(0)} \left( 1 - \a\L^2\frac{2}{27} \right). \nn
\eea
We can  also work out the black hole entropy \eqref{wald} giving
\be
S  = \frac{r_{H(0)}^2}{4 G} (1 + \frac{2 \alpha \Lambda^2}{27}).
\ee
Note that although both the temperature and the entropy are corrected at order $\alpha$ the combination
$T S $ is actually uncorrected at this order. Moreover, imposing the thermodynamic relation
\be
dM = T dS,
\ee
and using the fact that the $C^3$ term evaluated on pure AdS is zero, we can infer that the mass must also be unchanged at order $\alpha$. (In varying the entropy note that both $\alpha$ and $\Lambda$ are held fixed.) Using the relation $F = M - TS$ we can also then infer that the onshell action must also be unchanged at order $\alpha$.

One can also argue that the free energy and mass are unchanged at order $\alpha$ by considering their direct evaluation. Let us consider first the onshell action. The first step is to ensure that the variational problem, including the additional $C^3$ term, is well-posed at finite radius. To investigate this we vary the bulk action \eqref{bulk-weyl} with respect to the metric. From the term at order $\alpha$ one obtains the following contribution to the bulk field equation
\bea
{\cal G}^{\mu \nu} &=& \frac{1}{2} g^{\m\n}C_{\r\s}^{\ \ \t\eta}C_{\t\eta}^{\ \ \l\k}C_{\l\k}^{\ \ \r\s} \ -\  6C_{\r\s}^{\ \ \m \eta}C^\n_{\ \eta \l\k}C^{\l \k \r\s} \ +\  3 C^{\t\eta \l\k}C_{\l\k}^{\ \ \m \s}R^\n_{\ \s\t\eta}  \nn \\
&& +\  4 C^{\t\n \l\k}C_{\l\k}^{\ \ \m \s}R_{\t\s} \ -\  2C^{\t\n \l\k}C_{\l\k}^{\ \ \m \s}R g_{\t\s} \ -\  C^{\r\s\t\eta}C_{\r\s\t\eta}R^{\m\n} \nn \\
&& +\  6 \nabla_{\s} \nabla_{\t} (C^{\t\n \l\k}C_{\l\k}^{\ \ \m \s}) \ +\  2 \nabla_{\t} \nabla^{\m}(C^{\t\eta \l\k}C_{\l\k}^{\ \ \r\n}g_{\r\eta}) \ +\ \label{weyl1} \\
&& -\  \Box(2 C^{\m \eta \l\k}C_{\l\k}^{\ \ \r\n}g_{\r\eta} + C^{\r\s\t\eta}C_{\r\s\t\eta}g^{\m\n}) \ -\  2 \nabla_{\s} \nabla_{\t}(C^{\t\eta \l\k}C_{\l\k}^{\ \ \r\s}g_{\r\eta}g^{\m\n}) \nn \\
&&  +\  2\nabla_{\s} \nabla^{\m}(C^{\n \eta \l\k}C_{\l\k}^{\ \ \r\s}g_{\r\eta}) +\  \nabla^{\m} \nabla^{\n} (C^{\r\s\t\eta}C_{\r\s\t\eta}), \nn
\eea
where ${\cal G}_{\mu \nu} = R_{\mu \nu} - \frac{1}{2} R g_{\mu \nu} + \Lambda g_{\mu \nu}$. The variation results in the following boundary terms at order $\alpha$ involving  derivatives of the metric
\bea
&&\frac{1}{\k^2} \int_{\del {\cal M}} d^3x \ \sqrt{-\gamma} \ \a \left[ \ 3 n_\r C^{\r\s\eta \t}C_{\eta \t}^{\ \ \m\n}\nabla_\n \delta g_{\m\s} \ +\ n^\l C^{\r\s\eta \t}C_{\eta \t}^{\ \ \m\n}g_{\m\s}\nabla_\r \delta g_{\l\n} \right. \nn \\
&& \left. \ \ \ \ \ \ \ \ \ \ \ \ \ \ \ \ \ \  +\ n^\l C^{\r\s\eta \t}C_{\eta \t}^{\ \ \m\n}g_{\m\s}\nabla_\n \delta g_{\l\r} \ -\ n_{\z} C^{\r\s\eta \t}C_{\eta \t}^{\ \ \m\n}g_{\m\s}\nabla^{\z} \delta g_{\r\n} \right. \nn \\
&& \left. \ \ \ \ \ \ \ \ \ \ \ \ \ \ \ \ \ \   -\ n_\r C^{\r\s\eta \t}C_{\eta \t}^{\ \ \m\n}g_{\m\s}g^{\xi\l}\nabla_\n \delta g_{\xi\l} \ -\ C^{\m\n\r\s}C_{\m\n\r\s}\ \delta {\cal K} \  \right], 
\eea
with $n$ the normal to the boundary.
(There are additional boundary terms involving the metric which automatically vanish for a Dirichlet boundary condition.) As one would have anticipated, boundary terms involving the normal derivative of the metric arise in this variation and it would therefore seem as if one needs additional Gibbons-Hawking like terms in order for the variational problem to be well-posed. Moreover, working iteratively in $\alpha$ and then using the bulk field equations to simplify the boundary terms looks a very non-trivial calculation in this case. However, it turns out that one only needs to use the fact that the leading order metric is Einstein and is asymptotically locally AdS: in Fefferman-Graham coordinates \eqref{fg0}, the leading power in the Weyl tensor necessarily behaves as
\be
C_{\m\n\r\s} \sim \frac{1}{\rho^2}.
\ee
One can use this behaviour to show that the boundary terms needed for the variational problem to be well-posed all go to zero as a positive power of $\rho$. For example, the term
\be
\frac{1}{\kappa^2} \int_{\del {\cal M}} d^3x \sqrt{-\gamma} C^{\m\n\r\s} C_{\m\n\r\s} {\cal K}
\ee
evaluated at $\rho = \ep \ll1$ behaves as $\ep$, and therefore does not contribute in the limit  $\ep \rightarrow 0$. Therefore, although one could indeed use the explicit expansion of the onshell Weyl tensor to express the boundary terms in terms of quantities intrinsic to the boundary, the resulting boundary action cannot give a finite contribution to the onshell action.

The onshell action is thus given by
\bea
I_{\rm onshell} &=& \frac{1}{\kappa^2} \int d^4x \sqrt{-g} (\Lambda + \alpha C_{\m\n\r\s} C^{\r\s\eta\l} C_{\eta\l}^{\; \; \; \m\n}) \\
&& - \frac{1}{\kappa^2} \int d^3x \sqrt{-\gamma} \left ( K + 2/l + \frac{l}{2} R(\gamma) \right ), \nn
\eea
where we have used the onshell relation $R = 4 \Lambda + C^3$. The terms in the second line denote the Gibbons-Hawking term along with the counterterms. The latter suffice to remove the divergences at leading order, but do not in general suffice to remove additional divergences at order $\alpha$. However, it again turns out that the Weyl correction falls off sufficiently fast at the boundary that there are no additional terms needed at order $\alpha$. To show this one needs to use the fact that, in Fefferman-Graham coordinates, $C^3$ is of order $\rho^6$ or smaller and the correction to the metric at order $\alpha$ is of order $\rho^6$ or smaller. Looking at the terms in the onshell action, this means that the contributions at order $\alpha$ are of
order $\ep^3$ or smaller, and thus vanish in the limit $\ep \rightarrow 0$. For example, the term
\be
\int d^4 x \sqrt{-g} C^3 \sim \int_{\ep} d \rho \frac{1}{\rho^4} \cdot \rho^6 \sim \ep^3.
\ee
Thus only the above terms are needed in computing the renormalised action. Note that this implies that, as expected, the onshell action for AdS is uncorrected at order $\alpha$, regardless of the choice of conformal class of the boundary metric. In particular, if one computes the free energy for the dual theory on an $S^3$, it is not changed at order $\alpha$.

It is still non-trivial that the actual value of the free energy for the planar black hole is uncorrected, as the metric is corrected, the horizon position is shifted and the $C^3$ term in the action all give finite contributions at the horizon. Explicitly evaluating the onshell action using \eqref{aWeyl} together with
\be
C^3 = \frac{12 m^3}{r^9}
\ee
one obtains
\bea
F =  - \beta_T I^E_{\rm onshell} &=& \frac{V_{xy}}{\kappa^2} \int_{r_{h}}^{r_c} dr r^2 b(r) (\Lambda + \alpha \frac{12 m^3}{r^9} ) \\
&& + \frac{V_{xy}}{\kappa^2}( \sqrt{a(r)} \pa_r( \sqrt{a(r)} b(r) r^2) + \frac{2}{l} \sqrt{a(r)} b(r) r^2)_{r_c}, \nn
\eea
where $r_c \gg 1$ is the cutoff radius. Integrating the bulk term and looking at the horizon contribution one obtains
\be
\frac{V_{xy}}{\kappa^2}  (\frac{2}{3} \Lambda r^3 + \frac{4\alpha m^2 \Lambda}{r^3} - \frac{4 \alpha m^3}{r^6})_{r_{h}}  = \frac{V_{xy}}{\kappa^2} 2m.
\ee
i.e. the terms of order $\alpha$ cancel! Looking at the contribution from the cutoff boundary, as already argued the terms of order $\alpha$ fall off too quickly to contribute and one is left with a contribution
\be
- \frac{V_{xy}}{\kappa^2} \frac{5m}{2},
\ee
with the total free energy being
\be
F = - \frac{V_{xy}}{2 \kappa^2} m,
\ee
i.e. unchanged at order $\alpha$. One can similarly argue why the mass $M = V_{xy} m/\kappa^2$
is unchanged at this order: varying the renormalised onshell action with respect to the source for the stress energy tensor, all terms at order $\alpha$ are subleading in the radial expansion and do not contribute.

To summarise: the $C^3$ term leads to a correction of the metric of the planar black hole. The temperature and entropy are both changed at order $\alpha$ but the mass and the free energy are unchanged.

At this point we return to the physical interpretation of the integration constants in the corrected solution. The first integration constant $a_1$ corresponds to a shift in the mass parameter,
\be
m \rightarrow m - \alpha a_1.
\ee
This shift will affect the entropy, temperature, mass and free energy.
The second integration constant corresponds to a redefinition of the time coordinate and hence of the temperature. One can see this by looking at the form of the metric
\bea
ds^2 &=& - (1 + 2 \alpha b_1 + \cdots) a(r) dt^2 + \cdots \\
&=& - a(r) d \hat{t}^2 + \cdots \nn
\eea
i.e. by redefining the time coordinate one can absorb the integration constant $b_1$. This in turn corresponds to a shift of the temperature by
\be
T \rightarrow T (1 + \alpha b_1),
\ee
with the free energy and mass shifted by the same factor.

By an appropriate choice of the integration constant $a_1 = - m \Lambda^2/9$ one can make the entropy
be uncorrected at order $\alpha$. However, this value of $a_1$ is such that the temperature, mass and free energy are corrected:
\be
T \rightarrow T (1 - \frac{\alpha \Lambda^2}{9}); \qquad
M \rightarrow M (1 - \frac{\alpha \Lambda^2}{9}); \quad
F \rightarrow F (1 - \frac{\alpha \Lambda^2}{9}).
\ee
(These corrections are clearly consistent with the thermodynamic relation.) By fixing the integration constant $b_1$ appropriately and redefining the time coordinate, one can undo these corrections at order $\alpha$, leaving all thermodynamic quantities unchanged to order $\alpha^2$. However, such a redefinition is somewhat unnatural from the perspective of the holographic duality, as it implies that the time coordinate for the field theory is redefined at order $\alpha$.

Thus the thermodynamics at order $\alpha$ depends on which quantities one has chosen to hold fixed. In the context of supersymmetric black holes one fixes the mass (and charge), with the temperature necessarily being zero and the entropy being corrected by the higher derivative terms. In the context of finite temperature black holes, it would seem natural to fix the mass also, as we did above, with the temperature and the entropy being corrected.

\subsection{Einstein + $R^4$} \label{RiemannFour}

In order to obtain an eleven-dimensional correction which cannot be rendered trivial by field redefinitions we need to add a curvature invariant involving the Riemann tensor, with the first non-trivial term arising at fourth order. As discussed earlier, the reduction of this term will result in terms quartic in the Riemann tensor in the effective four dimensional action. In this section we will work with
one representative curvature invariant at this order, the same tensor structure considered earlier, but the generalization of the analysis to other tensor structures would be straightforward. The reason for considering this particular term is because, we discussed earlier, such a term would accompany any Riemann squared term occurring in the effective lower dimensional action.

The action we consider is therefore
\be \label{RiemannFourAction}
I = \frac{1}{2\kappa^2} \int d^4 x \  \sqrt{-g} \left ( R - 2 \L + \a (R_{\r\s\t\l} R^{\r\s\t\l})^2 \right )
\ee
The resulting field equation is
\bea
{\cal G}^{\m\n} &=& \frac{1}{2}(R_{\r\s\t\l}R^{\r\s\t\l})^2 g^{\m\n} \ -\  4R_{\r\s\t\l}R^{\r\s\t\l}R^{\m \b\g\delta}R^\n_{\ \b\g\delta} \  \\
&& +\  8\nabla_\r \nabla_\s(R_{\t\l\g\delta}R^{\t\l\g\delta}R^{\m \s\r \n}). \nn
\eea
Note also that there is a new boundary term involving derivatives of the metric variation obtained when varying the term at order $\alpha$
\be
\frac{4}{\k^2} \int_{\del {\cal M}} d^3x \ \sqrt{-\gamma} \ \a \  R_{\r\s\t\l}R^{\r\s\t\l}R^{\a\b\g\delta}\  n_{\g}
\nabla_\b \delta g_{\a\delta}.
\ee
We will discuss this term in the context of the variational problem below.

Now let us turn to the effect of such a correction on the planar black hole metric.
Evaluated on the static ansatz \eqref{DeserMetric} the action reduces to
\be
\hat{I} = \int_0^\infty dr\  \left[ 2rb^{\prime}(a-r^2) + \a \frac{D^2}{(br^2)^3} \right]
\ee
where to simplify formulae in this section we have imposed $\Lambda = - 3$ and
\bea
D &\equiv& 8a^2b'^2r^2+8a a'b'b r^2+4a'^2b^2r^2+4a^2b^2+9a'^2b'^2r^4+12a a'b'b''r^4\nn \\
&& +6a'a''b'b r^4 +4a^2b''^2r^4+4a a''b''b r^4+a''^2b^2r^4
\eea
Varying this action we obtain equations of motion for $a$ and $b$ consisting of the piece coming from the Einstein action and a piece proportional to $\a$ coming from the correction. Once again we solve these perturbatively in $\a$ as in Section \ref{f(R)Gravity}. The planar black hole solution is indeed corrected:
\bea
a(r) &=& a_{(0)}(r) + \a a_{(1)}(r) \label{aRiemann}
\\
       &=&r^2 - \frac{m}{r} + \a \left( \left( \frac{a_1}{r} -96 r^2 \right) -\frac{672 m^2}{r^4}-\frac{1200 m^3}{r^7}-\frac{536 m^4}{r^{10}} \right) \nn
       \eea
and
\bea
b(r) &=& b_{(0)}(r) + \a b_{(1)}(r)  \label{bRiemann} \\
       &=& 1 + \a \left( b_1+\frac{336 m^2}{r^6}+\frac{224 m^3}{r^9} \right). \nn
\eea
Again there are two arbitrary integration constants, corresponding to redefining the mass parameter and the time coordinate at order $\alpha$. These constants will be set to zero and their effect will be discussed further below. Note however that the AdS metric itself is corrected by the quartic Riemann term, since the latter does not evaluate to zero, unlike the previous Weyl example.

\subsubsection{Thermodynamics of corrected black hole solutions}

Let us first calculate the corrected horizon position and temperature of this black hole solution.
The horizon is located at
\be
r_H = \frac{m^{1/3}}{3} (1 + \a \frac{104}{3}),
\ee
and
the temperature of this black hole solution is given by:
\be
T = \frac{1}{4\pi}m^{1/3}(3-208\a)
\ee
We also work out the Wald entropy using \eqref{wald} giving
\be
S = \frac{A}{4 G} - \frac{4 \alpha}{\kappa^2} \int_H (R^{\mu\nu \rho \sigma}R_{\mu\nu \rho \sigma}) R^{vr vr} \sqrt{\gamma} d^2 x,
\ee
where we use ingoing coordinates for the horizon, namely
\be
ds^2 = - a(r) b(r)^2 dv^2 + 2 b(r) dr dv + r^2 (dx^2 + dy^2),
\ee
and the integral is over the spatial part of the horizon. The integrand in the second term however vanishes when evaluated on the leading order solution since
\be
R^{vr vr} = \frac{1}{2} a'' = 2 - \frac{2m}{r^3}
\ee
which vanishes at the horizon. (The Riemann squared term is non-zero at the horizon.)
Hence, in this case, the black hole entropy is given only in terms of the corrected area of the horizon
\be
S = \frac{V_{xy}}{4 G} m^{2/3} \left(1 + \a \frac{208}{3} \right).
\ee
Combining the entropy and the temperature we notice that the combination
\be
T S  = \frac{V_{xy}}{2 \kappa^2} m + {\cal O}(\alpha^2)
\ee
is again unchanged at order $\alpha$. As in the previous section we can now argue that for the relation
\be
dM = T dS
\ee
to hold the correction to $dM$ at order $\alpha$ must also vanish.

This argument on its own however does not exclude there being a term in the mass (and free energy) at order $\alpha$ which is independent of the parameter $m$: recall that, unlike the Weyl example, the AdS metric itself is corrected and the $R^4$ term in the action evaluated on AdS is non-zero. In other words, the holographically renormalized higher derivative action evaluated on AdS could be non-vanishing. Computation of the renormalised mass and action is somewhat involved as it requires analysing the corrections to asymptotically locally AdS solutions, isolating the divergences, computing the counterterms and so on. Fortunately there is a short cut: when the dual field theory is supersymmetric, the mass of the $m=0$ solution is necessarily zero, as is the free energy, and therefore there cannot be any contributions  to the mass and free energy at order $\alpha$ which are independent of $m$. (Note that the free energy of the dual theory on a curved space would in general indeed be expected to be corrected at order $\alpha$.)

Thus, in summary, as for the $C^3$ case, the temperature and the entropy are corrected whilst the mass and free energy are not. By choosing the integration constants $a_1$ and $b_1$ one can adjust which thermodynamic quantities are corrected at order $\alpha$, but the most natural physical choice from holographic considerations is indeed that where the mass is fixed and the entropy is corrected.

\subsection{Corrections arising in string theory}

In this section we have explored the effect of various curvature invariants added to the four dimensional action. We have shown that the thermodynamic properties of black hole solutions are in general corrected even when the metric is not corrected. From a top down perspective it would be complicated to determine which curvature invariants arise in the four dimensional action, with a given higher dimensional invariant contributing to invariants of different derivative order in four dimensions.

From the dual holographic perspective, one can try to restrict the invariants which arise in four dimensions using the free energy on an $S^3$. This would not restrict at all Weyl invariants which do not contribute to the free energy. One would also anticipate that other specific combinations of invariants involving Riemann, Ricci and Ricci scalar can be made in which the correction to the free energy also vanishes.

In the following section we will turn to another criterion for higher derivative corrections: the spectrum of fluctuations and the corresponding dual operators. In general, imposing that such corrections lead to CFT operators which are unitary and have positive norm, rules out many curvature invariants.

\section{Spectrum of curvature squared theories} \label{spectrum}

\subsection{Linearized equations of motion}

In this section we discuss the spectra of the higher derivative theories. For the sake of brevity, we will mostly focus on the case of curvature squared corrections, but the analysis for other higher-derivative theories would be similar and will be discussed at the end.

We consider again the action \eqref{GaussBonnetAction}, whose equations of motion are given in \eqref{GaussBonnetEom} to \eqref{emn}. Since our interest here is in the context of holography, we consider the spectrum of excitations around $AdS_4$, and when we need an explicit form for the metric 
we will work in the Poincar\'{e} patch in which
\be
ds^2 = \frac{d\r^2}{\r^2} + \frac{1}{\r^2}(-d t^2 + d x^2 + d y^2).
\ee
We denote by $x^i$ the coordinates on the three dimensional slices of constant $\r$.
We vary the metric as $g_{\m\n} \rightarrow  \hat{g}_{\m\n} + \delta g_{\m\n} = \hat{g}_{\m\n} + h_{\m\n}$, where $\hat{g}_{\m\n}$ is the $AdS_4$ background metric. It was shown in \cite{Pope1} that the linear variations of the various tensors appearing in the equations of motion are
\bea
{\cal G}^L_{\m\n}  &=& R^L_{\m\n} - \frac{1}{2}R^L \hat{g}_{\m\n} - \L h_{\m\n} \nn \\
R^L_{\m\n}            &=& \nabla^\l \nabla_{(\m} h_{\n)\l} - \frac{1}{2} \Box h_{\m\n} - \frac{1}{2} \nabla_\m \nabla_\n h \label{RicciLinear} \\
R^L                         &=&  \nabla^\m \nabla^\n h_{\m\n} - \Box h - \L h \nn
\eea
where $\nabla_{\m}$ is the covariant derivative associated with $\hat{g}$. Note that $h = \hat{g}^{\mu \nu} h_{\mu \nu}$. We write explicitly
\bea
R &=& (\hat{g}^{\m\n} - h^{\m\n}) R_{\mu \nu} \\
    &=& (\hat{g}^{\m\n} - h^{\m\n}) (R^{(0)}_{\m\n} + R^L_{\m\n}) \nn \\
    &\equiv& R_{(0)} + R^L, \nn
\eea
where $R_{(0)}$ is the Ricci scalar in the background.

To linear order in the variation, the equations of motion then become \cite{Pope1}
\bea \label{VariationEom}
\delta({\cal G}_{\m\n} + E_{\m\n}) &=& [ 1 + 2 \L(\a + 4\b) ]{\cal G}^{L}_{\m\n} + \a [ (\Box - \frac{2\L}{3}){\cal G}^{L}_{\m\n} - \frac{2\L}{3}R^{L} \hat{g}_{\m\n} ] \nn \\
&& + (\a+2\b)[-\nabla_{\m} \nabla_{\n} + \hat{g}_{\m\n}\Box + \L \hat{g}_{\m\n}]R^L.
\eea
The most common gauge used in holography is radial axial gauge, $h_{\r\m} = 0$ for $\m = (\r,t,x,y)$. However, \cite{Pope1} used a covariant gauge
\be \label{PopeGauge}
\nabla^\m h_{\m\n} = \nabla_\n h,
\ee
since in this gauge the equations of motion immediately simplify.
\\
\\
Substituting \eqref{PopeGauge} into the linearized tensors \eqref{RicciLinear} gives
\bea
{\cal G}^L_{\m\n} &=& \frac{1}{2}\nabla_\m\nabla_\n h- \frac{1}{2}\Box h_{\m\n}+\frac{\L}{3} h_{\m\n} +\frac{1}{6}\L h \hat{g}_{\m\n}  \\
R^L &=& -\L h \nn
\eea
which can then be substituted into \eqref{VariationEom}, the variation of the equations of motion. Tracing over the result yields
\be \label{traceeqn}
0 = \delta({\cal G}_{\m\n}+{\cal E}_{\m\n}) = \L[ h - 2(\a+3\b) \Box h ].
\ee
Imposing the above constraint equation for the trace, we find that the variation of the field equations is,
\bea \label{GaugeVariationEom}
0 \ =\  \delta({\cal G}_{\m\n}+{\cal E}_{\m\n}) &=& -\frac{\a}{2} \Box^2 h_{\m\n} -\frac{1}{2}\left( 1+\frac{2\L\a}{3}+8\L\b \right)\Box h_{\m\n} \nn \\
&& +\frac{\L}{3}\left( 1+\frac{4\L\a}{3}+8\L\b \right) h_{\m\n}  \\
&& +3(\a+2\b)\left( \frac{1}{4(\a+3\b)}+\L \right)\nabla_\m\nabla_\n h \nn \\
&&  -\frac{\L}{12}\left( \frac{5\a+6\b}{\a+3\b}+\frac{4\L}{3}(\a+6\b) \right) \hat{g}_{\m\n}h. \nn
\eea
In \cite{Pope1} this equation was analysed only in the case of $(\alpha + 3 \beta) = 0$, with a view to critical gravity, but we will not impose this constraint here.
From \eqref{GaugeVariationEom} we wish to extract the equation of motion for $h_{\langle\m\n\rangle}$, the traceless part of $h_{\m\n}$, where
\be \label{Tracelessh}
h_{\m\n} = h_{\langle\m\n\rangle} + \frac{h}{4} \hat{g}_{\m\n}.
\ee
This yields, provided that $\beta \neq 0$,
\bea \label{TracelessHEom}
0 &=& -\frac{\a}{2}\left( \Box - \frac{2\L}{3} \right) \left( \Box + \frac{1}{\a} + \frac{4\L}{3} + \frac{8\b\L}{\a} \right)h_{\langle\m\n\rangle} \\
&& + \frac{3}{4}\frac{(\a+2\b)}{(\a+3\b)} \left(1+4\L(\a+3\b) \right)\nabla_{\langle\m}\nabla_{\n\rangle}h. \nn
\eea
This equation represents an inhomogeneous equation for the traceless part of the metric fluctuation. However one can rewrite the equation as a homogeneous equation
by defining a new traceless tensor $\psi_{\langle\m\n\rangle}$ as
\be \label{defpsi}
\psi_{\langle\m\n\rangle}=h_{\langle\m\n\rangle} + \l \nabla_{\langle\m}\nabla_{\n\rangle}h,
\ee
and choosing $\l$ such that the final term in \eqref{TracelessHEom} is zero. This value turns out to be
\be \label{lambda}
\l = -\frac{6(\a+3\b)}{3+8\L(\a+3\b)},
\ee
making the resulting equation of motion for $\psi$ homogeneous
\be \label{eom}
0 = \left( \Box - \frac{2\L}{3} \right) \left( \Box  - \frac{2\L}{3} - M^2 \right)\psi_{\langle\m\n\rangle},
\ee
where
\be
M^2 = -2\L-\frac{1}{\a}-\frac{8\L\b}{\a}. \label{mass2}
\ee
Moreover it is easy to verify that $\psi_{\langle\m\n\rangle}$ is transverse.

In the case where $\alpha = 0$, the analogue of \eqref{TracelessHEom} is
\be
(1 + 8 \beta \Lambda) (\Box - \frac{2 \Lambda}{3}) h_{\langle \m \n \rangle} + (1 + 12 \beta \Lambda) \nabla_{\langle \m} \nabla_{\n \rangle} = 0,
\ee
which can be rewritten as a homogeneous equation
\bea
(\Box - \frac{2 \Lambda}{3}) \psi _{\langle \m \n \rangle} &=& 0; \\
\psi_{\langle\m\n\rangle}&=& h_{\langle\m\n\rangle} - \frac{6 \beta}{1 + 8 \beta \Lambda} \nabla_{\langle\m}\nabla_{\n\rangle}h. \nn
\eea
Note that this equation is only second order.

The interpretation of these equations is as follows. In the Einstein theory the only propagating mode is the traceless part of the metric,
which couples to the dual stress energy tensor. In the theory with generic values of $(\alpha, \beta)$ the trace of the metric is a propagating mode
dual to a scalar operator ${\cal O}_h$ of dimension
\be
\Delta_{{\cal O}_h} = \frac{3}{2} + \sqrt{\frac{9}{4} + \frac{1}{2 \alpha + 6 \beta}}
\ee
whilst the equation of motion for the traceless part of the metric fluctuation is fourth order. One can write a basis for solutions of this fourth order equation
as
\bea
\psi_{\langle\m\n\rangle} &=& \psi^{(1)}_{\langle\m\n\rangle} + \psi^{(2)}_{\langle\m\n\rangle}; \\
(\Box - \frac{2 \Lambda}{3}) \psi^{(1)}_{\langle \m \n \rangle} &=& 0; \nn \\
(\Box - \frac{2 \Lambda}{3} - M^2) \psi^{(2)}_{\langle \m \n \rangle} &=& 0, \nn
\eea
with the propagating massless mode $\psi^{(1)}_{\langle \m\n \rangle}$ coupling to the dual stress tensor and the new mode $\psi^{(2)}_{\langle \m \n \rangle}$ being associated with a spin two operator $X$ of dimension
\be
\Delta_{X} = \frac{3}{2} + \frac{1}{2} \sqrt{ 9 + M^2}.
\ee
In this section we will show explicitly how these modes are associated with the dual spin two operator and we will discuss how the variational problem is defined.

Note that the special case in which the action on $AdS$ is uncorrected, with the bulk term in the action reducing to Riemann squared,  is obtained by choosing $\alpha = 4 \gamma$, $\beta = - \gamma$, in which case
\bea
\Delta_{{\cal O}_h} &=& \frac{3}{2} + \sqrt{\frac{9}{4} + \frac{1}{2 \gamma}}; \label{r2} \\
\Delta_{X} &=&  \frac{3}{2} + \frac{1}{2} \sqrt{ 9 - \frac{1}{\gamma}}.  \nn
\eea
We will discuss later when these operators are unitary, but let us already note here that for $\gamma \rightarrow 0$, which is indeed the case when the higher derivative corrections are small, either one or the other operator necessarily has a complex dimension and thus violates unitarity.

For special values of $(\alpha, \beta)$ one has to look more carefully to obtain the spectrum. At $\alpha= 0$ the higher derivative term consists just of the Ricci scalar squared, and the only new propagating mode in the bulk is the trace of the metric fluctuation, dual to a scalar operator. At $\alpha + 3 \beta = 0$, when the bulk term reduces to Weyl squared, the metric trace is zero so there is no dual scalar operator but there is still a propagating spin two mode dual to a spin two operator. Whenever
\be
(1 + 2 \Lambda \alpha + 8 \Lambda \beta) = 0,
\ee
the second spin two mode becomes massless, with the dual operator becoming the logarithmic partner of the stress energy tensor in the dual (L)CFT \cite{Johansson:2012fs}. Note that this mode can become massless even when the trace is a propagating mode, with $\alpha + 3 \beta =0$ being an additional constraint used to remove the scalar operator.

\subsection{Derivation of equations of motion in general gauge}

In this subsection we derive an elegant form for the linearized equations of motion without imposing a gauge.
Taking the trace of the equations of motion \eqref{GaussBonnetEom} to \eqref{emn} one obtains
\be
(2 \alpha + 6 \beta) \Box R - 12 - R = 0,
\ee
where we use the explicit value of the cosmological constant together with the Bianchi
identity
\be
\nabla^{\mu} R_{\mu \nu} = \frac{1}{2} \nabla_{\nu} R.
\ee
Now letting $r = (R+12)$ one obtains a diagonal equation of motion for $r$
\be \label{trace1}
(2 \alpha + 6 \beta) \Box r - r = 0.
\ee
Note that this equation did not rely on the linearized approximation and is exact.
Defining
\be
R_{\mu \nu} + 3 g_{\mu \nu} = s_{\mu \nu} + \frac{1}{4} r g_{\mu \nu}
\ee
where $s_{\mu \nu}$ is traceless, i.e. $g^{\mu \nu} s_{\mu \nu} = 0$, the traceless part of the linearized equation of motion
gives
\be
(\alpha \Box + (1 - 4 \alpha - 24 \beta) ) s_{\mu \nu} = (\alpha + 2 \beta) \nabla_{\< \mu} \nabla_{\nu\>} r,
\ee
with the parentheses denoting the symmetric traceless combination. This equation can be diagonalized by defining
\be
\psi^r_{\mu \nu} = s_{\mu \nu} - \gamma \nabla_{\< \mu} \nabla_{\nu\>} r
\ee
with
\be
\gamma =  \frac{2\a + 3\b}{3(1-8\a-24\b)}
\ee
to give
\be \label{symtr1}
(\alpha \Box + (1 - 4 \alpha - 24 \beta) ) \psi^r_{\mu \nu} = 0.
\ee
The equations \eqref{trace1} and \eqref{symtr1} represent second order equations for the linearized curvature tensor and hold in any gauge.
In the covariant gauge $\nabla^{\mu} h_{\mu \nu} = \nabla_{\nu} h$ used previously
\be
r = 3 h; \qquad
s_{\mu \nu} = - \frac{1}{2} (\Box + 2)  h_{\< \mu \nu\>} + \frac{1}{2} \nabla_{\< \mu} \nabla_{ \nu \> } h.
\ee
A complete basis for the solutions to \eqref{trace1} and \eqref{symtr1} can be obtained by setting
the metric fluctuation $h_{\mu \nu}$ to be
\be
h_{\mu \nu} = h^{T}_{\mu \nu} + h^{X}_{\mu \nu},
\ee
with
\be
r(h^T) = 0; \qquad s_{\mu \nu} (h^T) = 0,
\ee
and $(r(h^X), \psi^r_{\mu \nu} (h^X))$ are non-zero, satisfying \eqref{trace1} and \eqref{symtr1}.

In understanding the holography dictionary it is useful to look at the asymptotic solutions for \eqref{trace1} and \eqref{symtr1}. Since $r$ is simply a scalar field, of a specific mass, the general asymptotic solution to \eqref{trace1} is as usual
\be
r (\rho,x) = \rho^{3 - \Delta_{{\cal O}_r}} (r_0(x) + \rho^2  r_2(x) + \cdots )
+ \rho^{\Delta_{{\cal O}_r} }(r_{2 \Delta -3}(x) + \cdots),  \label{curv}
\ee
where  $\Delta_{{\cal O}_r} = \frac{3}{2} + \sqrt{\frac{9}{4} + \frac{1}{2 \alpha + 6 \beta}}$. Here $r_0(x)$ acts as the source for the dual operator, with $r_{2 \Delta -3} (x)$ being the normalisable mode, and all other terms in the expansion being fixed by the field equation.

Equation \eqref{symtr1} is an equation for a massive spin two field of a given mass. Such fields are considered less frequently in holography (they were first analysed in \cite{Polishchuck}) but one can analyse the general asymptotic solutions to the field equations as follows. The independent solutions are
\bea
\psi_{\langle \r\r \rangle}(\r,\vec{k}) &=& \r^{d - \Delta} (f(x) + \cdots)  + \r^{\Delta} ( \tilde{f}(x) + \cdots) \nn \\
\psi_{\langle i \r \rangle}(\r,\vec{k}) &=& \r^{d - \Delta-1} (B_i(x) + \cdots) + \r^{\Delta - 1} (\tilde{B}_{i}(x) + \cdots)  \label{ips} \\
\psi_{\langle ij \rangle}(\r,\vec{k}) &=&  \r^{d - \Delta -2} (X_{ij}(x) + \cdots) + \r^{\Delta - 2} (\tilde{X}_{ij}(x) + \cdots), \nn
\eea
where
\be \label{gamma}
\Delta = \frac{d}{2} +\frac{1}{2}\sqrt{d^2+4M^2}
\ee
with $d=3$ in this case and $M^2$ given in \eqref{mass2}. The fields without tildes denote the non-normalizable modes and those with tildes are the normalizable modes. Only the transverse traceless part of $X_{ij}$ and $\tilde{X}_{ij}$ are independent data, however, since the field equations imply
\bea
X^{i}_{i} &=& \tilde{X}^{i}_{i} = 0; \\
B_{i} &=& - \frac{1}{2 - \Delta} \pa^{j} X_{ji}; \nn \\
\tilde{B}_{i} &=& - \frac{1}{\Delta -1} \pa^{j} \tilde{X}_{ij}; \nn \\
f &=& - \frac{1}{3 - \Delta} \pa^{i} B_{i}; \qquad
\tilde{f} = - \frac{1}{\Delta} \pa^i \tilde{B}_{i}. \nn
\eea
Thus the defining data for the spin two field indeed corresponds to a transverse traceless spin two operator in the dual field theory.

The new defining boundary data is $r_{(0)} (x)$ in \eqref{curv} and $X_{ij}$ in \eqref{ips}, namely the near boundary behaviour of the scalar curvature and the (trace adjusted) Ricci tensor.  One can obtain a geometric interpretation of these boundary conditions as follows. 
As commented earlier, the most natural gauge for holography is the radial axial gauge in which the metric perturbations satisfy $h_{\rho \mu} = 0$ and
\be
ds^2 = \frac{d \rho^2}{\rho^2} + \frac{1}{\rho^2} (\eta_{ij} + H_{ij}) dx^{i} dx^j.
\ee
In this gauge the linearized Ricci scalar $r[h]$ defined above is given by
\be
r [H] = \rho^2 \hat{R} [H] - \rho^2 {\rm tr} (H'') + 3 \rho {\rm tr} (H'),
\ee
with a prime denoting a radial derivative and $\hat{R}_{ij}$ being the linearized curvature of $H_{ij}$, namely
\bea
\hat{R}_{ij} &=& \frac{1}{2} \left ( \pa^k \pa_j H_{ik} + \pa^k \pa_i H_{jk} - \pa_i \pa_j {\rm tr}(H) - \pa^k \pa_k H_{ij} \right ); \nn \\
\hat{R} &=& \pa^i \pa_j H_{ij} - \Box {\rm tr}(H).
\eea
From the equation for the linearized Ricci scalar we note that the leading asymptotic behavior of the metric perturbation corresponding to the propagating scalar mode satisfies
\be
{\rm tr } H= \frac{1}{(3 - \Delta) (1 + \Delta)} \rho^{3-\Delta_{{\cal O}_r}} r_{(0)} + \cdots
\ee
We can also express this condition in a more geometric way, in terms of the extrinsic curvature of the hypersurface with induced metric $\gamma_{\mu \nu}$, as
\be
\gamma^{\mu \nu} {\cal L}_{n} K_{\mu \nu} \rightarrow - 4 r_{(0) x} \rho^{3-\Delta_{{\cal O}_r}}.
\ee 
Therefore the new data $r_{(0)}(x)$ supplied corresponds to specifying the boundary condition for the trace of the normal derivative of the extrinsic curvature.

\subsection{Two point functions}

To extract the two point functions of the scalar and spin two operators, the field equations alone do not suffice: one needs to compute the onshell renormalized action. This is a non-trivial issue, as even when the bulk field has a mass such that the dual operator would be unitary, the corresponding two point function of that operator is not guaranteed to be positive. In other words, the sickness of the higher derivative theory can manifest itself in negative norms.

A useful trick for obtaining the two point functions is the following, borrowed from the three dimensional discussions in \cite{Bergshoeff:2009aq}.
Let us first rewrite the bulk terms in the action as
\be
I =\frac{1}{2 \k^2} \int d^4x \ \sqrt{-g} \left[ R-2\L + (\beta + \frac{\a}{4}) R^2 + \alpha S^{\mu \nu} S_{\mu \nu} \right],
\ee
where $S_{\mu \nu}$ is the traceless part of the Ricci tensor. We next introduce a scalar auxiliary field $\phi$ and a traceless spin two auxiliary field $\phi^{\mu \nu}$ and write the action as
\bea
I &=& \frac{1}{2 \k^2} \int d^4x \ \sqrt{-g} \left[ R-2\L + (\b + \frac{\a}{4}) (2 R \phi - \phi^2) \right ] \\
&& + \frac{1}{2 \k^2} \int d^4x \ \sqrt{-g} \left[ \a (2 S^{\mu \nu} \phi_{\mu \nu} - \phi^{\mu \nu} \phi_{\mu \nu}) \right]. \nn
\eea
Eliminating the auxiliary fields using their equations of motion gives the previous action.
For this action, the boundary term needed for a well-defined variational problem is
\be
I_{GHY} = - \frac{1}{\k^2} \int d^3x \ \sqrt{-\gamma} \left ( \  K \left[ 1+( 2\b  + \frac{\a}{2}) \f  \right ] + \a K_{\mu \nu} \phi^{\mu \nu} \right ).
\ee
Note that the problems in setting up a variational problem have been solved here, by the introduction of the auxiliary fields. A similar approach to dealing with the variational problem in higher derivative theories was discussed in \cite{Hohm:2010jc}. 
The action with auxiliary fields admits Einstein manifolds as solutions, in which
\be
g_{\mu \nu} = \hat{g}_{\mu \nu}; \qquad
\phi = R = 4 \Lambda; \qquad \phi_{\mu \nu} = S_{\mu \nu} = 0.
\ee
The action of course also admits other solutions, but in this section we are interested in the spectrum around an Einstein solution. For such solutions, the boundary counterterms given previously in \eqref{bn1} renormalise the onshell action. Note that, as previously anticipated, when one looks at the leading order Einstein solutions, the boundary conditions for the auxiliary fields do not involve non-trivial data (i.e. unlike the metric boundary condition, the boundary data for the auxiliary fields is not specified by  arbitrary scalars or tensors)  and indeed this remains true when evaluating corrections on such solutions. When we compute the spectrum below, however, we find that there is indeed non-trivial data required for the auxiliary fields, which is expressed in terms of arbitrary scalars and tensors. 

Let us now consider perturbations around such an Einstein solution $\hat{g}_{\mu \nu}$ of the equations of motion, i.e.  we let
\be
g_{\mu \nu} = \hat{g}_{\mu \nu} + h_{\mu \nu}; \qquad
\f = 4\L + \delta \f; \qquad \phi_{\mu \nu} = \delta \phi_{\mu \nu}.
\ee
The boundary data for $\delta \phi$ and $\delta \phi_{\mu \nu}$ specify the defining data for dual scalar and tensor operators, respectively. 

Let us begin with the $\a = 0$ case.
To quadratic order in the fluctuations one obtains the following for the bulk terms in the action
\bea
\delta I
&=& -  \frac{\mu}{2 \kappa^2} \int d^4x\  \sqrt{-\hat{g}}\  h^{\mu \nu} \left( {\cal G}^L_{\m \n}[h] \right)  \\
&& + \frac{\b}{\kappa^2} \int d^4 x \sqrt{-\hat{g}} h^{\mu \nu} \left ( \nabla_{\mu} \nabla_{\nu} \delta \phi - \Box \delta \phi \hat{g}_{\mu \nu} \right ) \nn \\
&& + \frac{\b}{\kappa^2} \int d^4x \ \sqrt{-\hat{g}} \ \delta \f \left( R[h] - \Lambda h - \delta \f \right). \nn
\eea
where $\mu = (1 + 8 \b \Lambda)$, $h = \hat{g}^{\mu \nu} h_{\mu \nu}$ and the linearisation of the Einstein equation is
\be
{\cal G}^L_{\m\n} [h] =  R_{\m\n}[h] - \L h_{\m\n} - \frac{1}{2}R[h] \hat{g}_{\m\n} + \frac{1}{2 }\Lambda h \hat{g}_{\mu \nu},
\ee
and the linearised Ricci tensor is given by
\be
R_{\mu \nu} [h] = \frac{1}{2} \left ( \nabla^{\rho} \nabla_{\mu} h_{\rho \nu} + \nabla^{\rho} \nabla_{\nu} h_{\rho \mu} - \nabla_{\mu} \nabla_{\nu} h - \Box  h_{\mu \nu} \right ),
\ee
with $R[h] = \hat{g}^{\mu \nu} R_{\mu \nu}[h]$ being
\be
R[h] = \nabla^{\mu} \nabla^{\nu} h_{\mu \nu} - \Box h.
\ee
The action can be diagonalised with the field redefinition
\be
h_{\m\n} = \bar{h}_{\m\n} + \zeta(\delta \f)\hat{g}_{\m\n}
\ee
and letting
\be
\zeta = - 2\b/\mu.
\ee
The bulk action at the quadratic level then becomes
\bea
\delta I &=& - \frac{\mu}{2 \kappa^2} \int \ d^4x\  \sqrt{-\hat{g}} \bar{h}^{\m\n} \left( {\cal G}^L_{\m\n}[\bar{h}] \right) \\
&& - \frac{\b}{\kappa^2} \int d^4 x \sqrt{-\hat{g}} \delta \f \left[ - 6  \frac{\b}{\m} \Box \delta \f + \delta \f \right]. \nn
\eea
where we have used
\be
R(\delta \f \hat{g}_{\mu \nu}) = - 3 \Box \delta \f.
\ee
The equations of motion resulting from this action describe the graviton together with the scalar field, and agree with those found in the previous sections.

Having obtained the action, it is now straightforward to extract the two point functions of the dual operators. To do this we need to keep careful track of the boundary terms, include those which arise in the field redefinitions.
In working out these terms it is convenient to fix a gauge for the metric perturbation $\bar{h}_{\mu \nu}$, the holographic radial axial gauge in which
\be
\bar{h}_{\rho \mu} = 0.
\ee
Thus the perturbed metric may be written as
\be
ds^2 = \frac{d \rho^2}{\rho^2} + \frac{1}{\rho^2} \left ( \eta_{ij} + H_{ij} \right ) dx^i dx^j
\ee
where $H_{ij} = \rho^2 \bar{h}_{ij}$. Evaluating all boundary terms involving $H_{ij}$, including those from the counterterms needed to renormalise the action for the background solution, one obtains
\be
I_{\rm onshell} = - \frac{\mu}{4 \kappa^2} \int d^3x \frac{1}{\rho^2} \left ( H^{ij} \pa_{\rho} H_{ij} + 2 H \pa_{\rho} H \right ),
\ee
where $H = \eta^{ij} H_{ij}$. These terms can be processed using the Fefferman-Graham expansion, namely
\be
H_{ij} = H_{(0)ij} + \rho^3 H_{(3)ij} + \cdots
\ee
with $H_{(3)ij}$ being traceless and transverse. Thus
\be
I_{\rm onshell} = - \frac{3 \mu}{4 \kappa^2} \int d^3 x H^{ij}_{(0)} H_{(3) ij}.
\ee
This action is clearly finite, without the need for additional counterterms, as expected as the counterterms should take of all divergences of Einstein solutions of the field equations. Moreover, recalling that
\be
\< T_{ij} \> = - \frac{2}{\sqrt{g_{(0)}}} \frac{\delta I_{\rm onshell}}{\delta g^{(0)ij}}
\ee
we recover the formula
\be
\< T_{ij} \> = \frac{3 \mu}{2 \kappa^2} H_{(3) ij},
\ee
which is the linearisation of the renormalised stress tensor given earlier. Relative to Einstein gravity, this formula is shifted
by a factor of $\mu$ which in turn implies that the two point function for the stress energy tensor will be shifted by factor of $\mu$ relative to the Einstein case \cite{de Haro:2000xn}.
In this theory the ratio $\eta/s$ is unchanged by the higher order correction. This was already apparent on general grounds, since the correction evaluated on an Einstein solution can be removed by field redefinitions.
Here the derivation is somewhat non-trivial as both quantities are shifted by the factor of $\mu$: the Wald entropy was computed earlier, and $\eta$ is obtained from the two point function of the stress energy tensor, which according to the formula above will only be shifted by $\mu$ relative to the Einstein case.

What remains is to collect together all of the terms involving the scalar field. These give
\be
I_{\rm onshell} =  - \frac{13 \beta^2}{\mu \kappa^2} \int d \Sigma^{\mu} \pa_{\mu} (\delta \phi) \delta \phi. \label{end1}
 \ee
This is the action in Lorentzian signature. The corresponding action in Euclidean signature is
\bea
I^E_{\rm onshell} &=&  \frac{13 \beta^2}{\mu \kappa^2} \int d \Sigma^{\mu} \pa_{\mu} (\delta \phi) \delta \phi; \label{end2} \\
&=&  \frac{13 \beta^2}{\mu \kappa^2} \int d^3 x \frac{1}{\rho^2} \delta \phi \pa_{\rho} \delta \phi. \nn
\eea
Recalling that the asymptotic expansion of such a scalar field dual to an operator of dimension $\Delta$ is
\be
\delta \phi = \rho^{d- \Delta} (\delta \phi_{(d- \Delta)} + \dots) + \rho^{\Delta} (\delta \phi_{\Delta} + \cdots)
\ee
we see that this part of the onshell action still has divergences as $\rho \rightarrow 0$. This was indeed to be expected, as the counterterms computed earlier were for Einstein solutions of the field equations only.

The holographic renormalization required for such a scalar field is already known:
if the onshell (non-renormalized) Euclidean action for a free scalar field is
\be
I^E_{\rm onshell} = - \frac{1}{2} \int d \Sigma^{\mu} \varphi \pa_{\mu} \varphi
\ee
then the renormalised two point function of the operator of dimension $\Delta$ dual to the field $\varphi$ is
\cite{Skenderis:2002wp}
\be
\< {\cal O}_{\varphi} (x) {\cal O}_{\varphi} (0) \> = \frac{(2 \Delta -d) \Gamma(\Delta) }{\pi^{d/2} \Gamma(\Delta - d/2)} {\cal R} \left ( \frac{1}{x^{2 \Delta}} \right ) \equiv  c_{\Delta} {\cal R} \left ( \frac{1}{x^{2 \Delta}} \right ),
\ee
where we denote by ${\cal R}$ the renormalised quantity.
Comparing with our case we obtain the following for the norm of the two point function of the operator dual to $\delta \phi$:
\be
\< {\cal O}_{\phi} (x) {\cal O}_{\phi} (0) \> = - \frac{13 \beta^2}{\mu \kappa^2}
c_{\Delta} {\cal R} \left ( \frac{1}{x^{2 \Delta}} \right ).
\ee
The norm is never positive and recall that the operator also has complex dimension for negative $\beta$. It would be interesting to carry out a similar analysis for the
spin two operator, setting $\alpha \neq 0$, to find for which values of $\alpha$ the norm of the dual operator is non-positive. 

\subsection{Spectra for other curvature corrections}

A similar analysis can be carried out for the spectrum around AdS induced by other curvature corrections. For any given curvature invariant one might anticipate that generically the operator associated with the higher derivative term is either non-unitary or has non-positive norm. There are certain exceptions to this generic case, however.

If a curvature invariant which is built out of the Weyl tensor is added to the action, then the equations of motion linearised around AdS are necessarily unchanged since the Weyl tensor vanishes identically on the background. More precisely, any curvature at least cubic in the Weyl tensor implies that all contributions to the linearised field equations are at least linear in the Weyl tensor of the background, which vanishes for AdS. Therefore the Weyl terms do not change the spectrum of operators in the dual CFT, although they can modify the correlation functions of these operators. This fits with the observation that the Weyl terms on asymptotically locally AdS spacetimes fall off sufficiently fast at infinity that the variational problem is unchanged from the Einstein case. Put differently, one needs no additional new boundary conditions and therefore there are no new associated propagating modes and corresponding dual operators.

If one adds several different curvature invariants to the action, the diagonalization of the linearised field equations becomes more complicated, as we saw in the case of curvature squared corrections. For each new boundary condition there is an associated new dual operator. The dimensions and norms of the dual operators are obtained non-trivially from diagonalising the field equations and manipulating the onshell action.

From a top-down perspective, the leading higher derivative terms in the four-dimensional action must be consistent with unitarity. This implies that they must give rise to a linearised spectrum around AdS which is consistent with dual operators of real conformal dimension and positive norm. We have shown that individual terms such as $R^2$ are not consistent with unitarity, but we also noted that in reducing a higher dimensional curvature invariant any such curvature squared term always appears with fourth order curvature terms and a shift of the cosmological constant. Moreover, one needs to include all higher dimensional curvature invariants to respect unitarity at the required order, supersymmetry and so on.

Finally let us consider how the dimensions of the dual operators relate to the parameters of the dual CFT. The dimensions of the operators for the case of Riemann squared were given in \eqref{r2}. From the discussion around \eqref{mixing} we note that when such a term arises from a reduction of eleven dimensions the coupling constant $\gamma$ would be the ratio of the term descending from $R^4$ to the leading order Einstein term. In other words, for the case of an $S^7$ reduction $\gamma$ would be of order $1/N$ and in ABJM it would be of order $1/ (k N) = 1/ N'$. The dimensions of the operators scale as $1/\sqrt{\gamma}$, i.e. $N^{1/2}$ and $(N')^{1/2} = k \lambda^{1/2}$ respectively. Riemann squared on its own is not unitary, so one of the operators always has complex dimensions, but the combination with other terms arising from top down would give operators whose dimensions scale similarly. If no such operators exist in the dual CFT, then the net effect of the reduction of the leading top down terms must be trivial at the linearized level.

\section{Conclusions} \label{conc}

In this paper we showed that the variational problem is not in general well-posed in higher derivative gravity theories without specifying additional data to the boundary metric. When the higher derivative terms are treated perturbatively around the leading order Einstein solution, the higher derivative equations always become inhomogeneous second order equations, for which the variational problem is well-posed with only a boundary condition for the metric. However, in analyzing the spectrum around the corrected background, the linearized equations of motion are generically higher order and do indeed require additional boundary conditions. In the context of holography these additional boundary conditions correspond to data for operators in the dual conformal field theory. For the curvature invariants we analyzed the operators are non-unitary since their conformal dimensions are generically complex and their norms are non-positive definite. From a top down perspective, the reduction of any given higher dimensional curvature invariant results in a lower dimensional action involving several curvature invariants of different derivative order. When the lower dimensional curvature invariants are combined, the resulting spectrum must be unitary and thus either the dual operators must have real dimensions and positive norms or (perhaps more likely) the resulting lower dimensional linearized field equations remain second order with no new operators arising. 

Even when the new operators induced by the higher derivative terms are non-unitary, one might try to look for a unitary subsector of the theory, by switching off these operators. This is indeed the perspective of \cite{ChiralTMG, Maldacena}, whose boundary conditions effectively set to zero the sources for the irrelevant operators associated with the higher derivative terms. As explained in detail in \cite{Skenderis:2009nt,Skenderis:2009kd}, however, switching off such sources does not switch off expectation values for such operators. Moreover, even if one restricts to a subsector of the theory in which the irrelevant operators are neither sourced nor acquire expectation values, the theory itself is non-unitary. In particular, the extra fields do contribute in computation of stress energy tensor correlation functions and there is no guarantee that the latter would be unitary, nor is it immediately apparent that the additional operators can always be decoupled from the stress energy tensor. 

Many interesting questions deserve further study. By comparison with dual field theory results, one could, at least in the highly supersymmetric examples of ABJM and ABJ models, restrict what curvature invariants can arise in the effective four-dimensional action. It would also be interesting to work out the spectrum for the reduction of a top down curvature invariant, i.e. putting together curvature invariants of different order in four dimensions.  In this work we have found that the simplest representative correction would involve the Weyl tensor at orders higher than two: such corrections are in some ways analogous to the Gauss-Bonnet examples in higher dimensions, in that the black holes are corrected but there are no new operators induced in the spectrum. One of the initial motivations for looking at higher derivative terms in $AdS_4$ was to explore subleading effects in applied holography and the Weyl solution would be a natural candidate for such investigations. 

One approach to holographic cosmology exploits the domain wall cosmology correspondence 
\cite{Skenderis:2006jq}
to obtain a field theoretic description of cosmologies in one higher bulk dimension \cite{McFadden:2009fg}. Since the primary focus is naturally on four dimensional cosmologies, the results obtained here would be relevant in discussing higher derivative effects in holographic cosmology. In particular it would be interesting to understand whether corrections which give rise to non-unitary effects on the AdS side are automatically excluded from being physical on the cosmological side, and whether the irrelevant operators associated with the higher derivative terms could actually be useful on the cosmology side in, for example, exiting from the inflationary era.

\section*{Acknowledgements}

This work is part of the research program of the Stichting voor Fundamenteel Onderzoek der Materie (FOM), which is financially supported by the Nederlandse Organisatie voor Wetenschappelijk Onderzoek (NWO). JS acknowledges support via an NWO Vici grant. MT acknowledges support from a grant of the John Templeton Foundation. The opinions expressed in this publication are those of the authors and do not necessarily reflect the views of the John Templeton Foundation.

\end{document}